%% file: main.tex
\documentclass[lettersize,journal]{IEEEtran}

\usepackage{amssymb}
\usepackage{algorithmicx}

\usepackage{lineno}
\usepackage{subfig}
\usepackage{booktabs,caption}
\usepackage[flushleft]{threeparttable}
\usepackage{amsfonts}
\usepackage{float}
\usepackage{placeins}
\usepackage{textcomp}
\usepackage{amsmath}
\usepackage{graphicx}
\usepackage{url}
\usepackage{multicol, blindtext}
\usepackage{booktabs}
\usepackage[table]{xcolor}
\usepackage{xcolor}
\usepackage{algorithmicx}
\usepackage{lipsum}
\usepackage{amssymb}
\usepackage{lineno}
\usepackage{algpseudocode}
\usepackage{amsfonts}
\usepackage{siunitx}
\usepackage{amsmath}
\usepackage{multirow}
\usepackage{wrapfig}
\usepackage{array}
\usepackage{color}
\usepackage{silence}
\usepackage[table]{xcolor}
\definecolor{lightgray}{gray}{0.9}
\usepackage{subcaption}
\usepackage[linesnumbered,ruled,vlined]{algorithm2e}

\usepackage{tabularx}

\usepackage{amssymb}
\usepackage{lineno}
\usepackage{subfig}
\usepackage[english]{babel}
\usepackage{ragged2e}
\usepackage{blindtext}
\usepackage{booktabs,caption}
\usepackage[flushleft]{threeparttable}
\usepackage{amsfonts}
\usepackage{float}
\usepackage{placeins}
\usepackage{textcomp}
\usepackage{amsmath}
\usepackage{graphicx}
\usepackage{url}
\usepackage{multicol, blindtext}
\usepackage{booktabs}
\usepackage[table]{xcolor}
\usepackage{xcolor}
\usepackage{lipsum}
\usepackage{amssymb}
\usepackage{lineno}
\usepackage{algpseudocode}
\usepackage{amsfonts}
\usepackage{siunitx}
\usepackage{amsmath}
\usepackage{multirow}
\usepackage{wrapfig}
\usepackage{array}
\usepackage{color}
\usepackage{silence}
\usepackage[table]{xcolor}
\definecolor{lightgray}{gray}{0.9}
\usepackage{svg}

\begin{document}

\title{A Lightweight Defense Mechanism against Next Generation of Phishing Emails using Distilled Attention-Augmented BiLSTM}

\author{Morteza Eskandarian, Mahdi Rabbani, Arun Kaniyamattam, Fatemeh Nejati, Mansur Mirani, Gunjan Piya, Igor Opushnyev, Ali A. Ghorbani, ~\IEEEmembership{Senior Member, IEEE}, Sajjad Dadkhah,~\IEEEmembership{Senior Member, IEEE}
\thanks{M. Eskandarian, M. Rabbani, A. Kaniyamattam, F. Nejati, A A. Ghorbani, and S. Dadkhah are with the Canadian Institute for Cybersecurity, Faculty of Computer Science, University of New Brunswick, Fredericton, NB, Canada (e-mails: \{morteza.eskandarian, m.rabbani, arun.kaniyamattam, fatemeh.nejati, ghorbani, sdadkhah\}@unb.ca). M. Mirani, G. Piya, I. Opushnyev are with the Mastercard Vancouver Tech Hub, Vancouver, British Columbia, Canada (e-mails: \{mansur.mirani, sunny.piya, igor.opushnyev\}@mastercard.com).}}

\markboth{}
{Shell \MakeLowercase{\textit{et al.}}: A Sample Article Using IEEEtran.cls for IEEE Journals}


\maketitle

\begin{abstract}
\input{abstract}
\end{abstract}


\section{Introduction}
\input{Introduction}


\section{{Related Work}}
\input{relatedwork}
\section{{Data Augmentation and Synthetic Data Generation}}
\input{Data_agmentation}

\section{{Methodology}}

\input{methodology}

\section{Experimental results and evaluation}
\input{evaluation}

\section{Conclusion}
\input{conclusion}
\section*{Acknowledgements}

\input{Acknowledgments.tex}

\bibliographystyle{IEEEtran}
\bibliography{References}


\end{document}

%% file: abstract.tex
The current generation of large language models produces sophisticated social-engineering content that bypasses standard text screening systems in business communication platforms. Our proposed solution for mail gateway and endpoint deception detection operates in a privacy-protective manner while handling the performance requirements of network and mobile security systems. The MobileBERT teacher receives fine-tuning before its transformation into a BiLSTM model with multi-head attention which maintains semantic discrimination only with 4.5 million parameters. The hybrid dataset contains human-written messages together with LLM-generated paraphrases that use masking techniques and personalization methods to enhance modern attack resistance. The evaluation system uses five testing protocols which include human-only and LLM-only tests and two cross-distribution transfer tests and a production-like mixed traffic test to assess performance in native environments and across different distribution types and combined traffic scenarios. The distilled model maintains a weighted-F1 score difference of 1-2.5 points compared to the mixture split results of strong transformer baselines including ModernBERT, DeBERTaV3-base, T5-base, DeepSeek-R1 Distill Qwen-1.5B and Phi-4 mini while achieving 80-95\% faster inference times and 95-99\% smaller model sizes. The system demonstrates excellent performance in terms of accuracy and latency while maintaining a compact size which enables real-time filtering without acceleration hardware and supports policy-based management. The paper examines system performance under high traffic conditions and security measures for privacy protection and implementation methods for operational deployment. The research team will release all necessary code and training scripts and corpus splits to support security researchers who want to reproduce experiments and implement practical solutions.

%% file: Introduction.tex
\label{Introduction} 

The core operations of enterprise workflows depend on application-layer communication through email systems and messaging and collaboration platforms. The main source of social engineering attacks and credential theft is user-generated content because attackers use human vulnerabilities and business environment details to outmaneuver security boundaries. The combination of content filters with URL and sender reputation signals proves effective against established threat indicators yet becomes ineffective when attackers use diverse wording, structural methods, and contextual variations at large scales  \cite{alkhalil2021phishing, chiew2018survey, ENISA2023ThreatLandscape}. The emergence of large language models (LLMs) has made this problem worse by generating sophisticated messages that duplicate authentic business communication styles and legitimate tones which evade traditional keyword detection systems and fixed security protocols  \cite{nasution2025benchmarking,motlagh2024large, nahmias2024prompted}.Research indicates that generative and paraphrased text can achieve higher evasion rates than typical attacks, underscoring the need for LLM-aware defenses \cite{ebrahimi2017hotflip}.

Security detectors need to achieve two main goals: they must resist changes in content distribution between human and LLM-generated data while operating effectively under memory, latency, and privacy constraints at gateways and endpoints. Security applications and policy-driven defenses face the challenge of meeting operational constraints while preserving semantic discrimination.

The research develops a lightweight bidirectional LSTM model with multi-head attention that maintains both deployment readiness and robustness through distillation from a MobileBERT teacher model. The established edge and mobile strategy of knowledge distillation enables the transfer of contextual decision boundaries from transformer models to efficient student models \cite{hinton2015distilling, gou2021knowledge, sun2020mobilebert}. The attention-augmented BiLSTM model processes local phrase information and global semantics while maintaining low memory usage and stable processing times, which makes it suitable for enterprise communication pipeline integration.

The hybrid dataset contains both public human-generated content and LLM-produced variants of this content to simulate modern adversarial attacks against detectors. The synthetic examples contain paraphrasing, masking, and personalization techniques which have been shown to enhance evasion capabilities and test systems that depend on fixed indicators \cite{guo2024x, ebrahimi2017hotflip}. The data design follows evidence showing systems perform better when they receive inputs from diverse sources and undergo controlled perturbations \cite{wang2022generalizing, zhou2022domain}. The evaluation assesses the distilled student model against multiple transformer models which operate in latency-critical NLP applications including ModernBERT, DeBERTaV3-base, T5-base, DeepSeek-R1 Distill Qwen-1.5B, Phi-4 mini, and the MobileBERT teacher \cite{warner2024smarter, he2020deberta, he2021debertav3, guo2025deepseek, raffel2020exploring, abouelenin2025phi}. The research evaluates the system through measurements of accuracy and weighted F1 scores and inference time and parameter counts across GPU and mobile-class hardware platforms to simulate real-world network and mobile security environments.

This work presents four main contributions. 
\begin{itemize}
    \item \textbf{LLM-aware evaluation corpus.} A hybrid dataset combines public data sources with LLM-generated paraphrases that include masking and personalization features to create a realistic test environment for application-layer defense systems in enterprise communication networks.
    \item \textbf{Deployable distilled architecture.} The proposed attention-augmented BiLSTM model uses a transformer teacher for training while maintaining semantic discrimination and achieving small memory usage and predictable low-latency processing for gateway and endpoint deployment.
    \item \textbf{Accuracy-latency-size trade-off analysis.} The evaluation compares our model against current transformer models to determine optimal settings for real-time filtering operations. The student model maintains a performance level within approximately 1-2.5 weighted-F1 points of leading transformer models while achieving significant reductions in both inference time and parameter count.
    \item \textbf{Operational perspective.} The paper examines system performance under high traffic conditions, explains how to implement on-device filtering for privacy protection, and describes how to integrate the method into mail systems and client applications while following information security protocols.
\end{itemize}

In summary, the combination of strategic knowledge transfer with adversary-aware data augmentation produces detectors that protect against LLM-assisted deception attacks while meeting the operational needs of enterprise information security applications.

The paper continues with the following structure. The current paper evaluates existing research about phishing email detection methods in Section used LLMs to create artificial emails for analysis. The paper describes the proposed methodology beginning with the \ref{relatedWork}. The paper explains how researchers obtained phishing and legitimate email samples and how they baseline model and then shows how attention mechanisms and knowledge distillation improve the model. The evaluation section of this paper presents experimental results to show how each model performs. The paper concludes in Section \ref{conclusion} by summarizing the research outcomes and identifying possible research paths for the future.


%% file: relatedwork.tex
\label{relatedWork}

The growing frequency and sophistication of phishing attacks in recent years have underscored the need for more effective detection strategies \cite{thakur2023systematic}. A critical factor in developing robust models is the availability of high-quality, representative datasets. Traditional phishing datasets have played a key role in training machine learning models by capturing structural and lexical differences between benign and malicious communications. However, the emergence of Large Language Models (LLMs) has introduced a new challenge, as adversaries now exploit these models to generate highly realistic phishing content \cite{heiding2024devising}. As a result, models trained only on conventional data often struggle to detect these advanced threats. Recent studies have investigated both the use of LLMs to generate phishing emails and their role in detection, underscoring the need for LLM-aware models and datasets. The following review focuses on recent efforts that leverage LLMs for phishing detection, evaluate LLM-based phishing risks, or propose new datasets to address this evolving threat landscape.

Nahmias et al. \cite{nahmias2024prompted} proposed a document vectorization approach that utilizes multiple LLMs to detect spear-phishing emails. Their method introduces \textit{prompted contextual document vectors}, which are generated by querying several LLMs with human-crafted questions based on persuasive principles commonly associated with social engineering. Using a unique dataset of 333 LLM-generated spear-phishing emails, the model achieved an F1 score of 91\%, despite being trained only on classic phishing and benign emails. The approach demonstrated good performance over traditional document embedding techniques, achieving a 96\% recall with a simple k-nearest neighbors classifier. However, the method faces limitations due to the manual effort required in designing effective prompts and the high computational overhead associated with using large-scale models (exceeding 175 billion parameters). Additionally, the evaluation primarily focused on LLM-generated spear-phishing emails, leaving questions about its effectiveness on human-crafted attacks.


Koide et al. \cite{koide2024chatspamdetector} introduced ChatSpamDetector, a system that leverages LLMs to classify spam emails by transforming both email body content and header information into LLM-compatible prompts. Utilizing GPT-4, the system achieved higher precision across multilingual spam detection tasks. One of its key strengths is the ability to generate detailed justifications for its classifications, allowing users to make more informed decisions regarding suspected spam. Despite its high accuracy, the system faces challenges due to the computational cost and dependency on external API access. When evaluated on a dataset containing both spam and genuine emails, ChatSpamDetector outperformed traditional machine learning-based spam filters, which typically achieve precision rates between 54\% and 86\%. The system also demonstrated strong effectiveness in identifying complex social engineering tactics and brand impersonation attempts, even when basic email security features had been compromised.

Bethany et al. \cite{heiding2024devising} conducted a large-scale, 11-month study on LLM-based lateral spear phishing attacks, targeting approximately 9,000 staff members. Their findings revealed that emails generated using LLMs were as effective as those crafted by humans, with nearly 10\% of recipients disclosing their login credentials. The study also highlighted that emails incorporating internal organizational information were significantly more successful than those relying on external content, yielding entry rates of 9.76\% versus 0.21\%, respectively. Notably, individuals previously compromised in past campaigns were found to be 2-3 times more susceptible to LLM-generated phishing emails, even after participating in awareness training. To address these challenges, the authors proposed T5-LLM Email Defense, a detection model that achieved a 98.96\% success rate in identifying LLM-generated phishing emails. The study underscored critical weaknesses in existing anti-phishing systems and emphasized the urgent need for defense mechanisms specifically designed to counter the evolving threat posed by LLM-generated content.

Hazell \cite{hazell2023spear} conducted a comprehensive investigation into the use of LLMs in spear phishing campaigns, analyzing their applicability across various stages of the cyber kill chain, with particular focus on reconnaissance and message generation. Using GPT-3.5 and GPT-4, the study generated over 600 unique spear phishing emails targeting British Members of Parliament, demonstrating high levels of personalization at minimal cost-less than a fraction of a cent per message. The results revealed that basic prompt engineering techniques could bypass the embedded safety constraints of commercial LLMs, exposing substantial security risks. The generated emails effectively combined personal details, contextual relevance, and psychological manipulation, illustrating the capacity of modern LLMs to craft persuasive and targeted phishing content. While the study highlighted the emerging threat of AI-powered phishing, it also proposed mitigation strategies, including structured access controls and LLM-specific defense mechanisms. A noted limitation was the study's narrow focus on email-based attacks, acknowledging that future threats may span diverse communication platforms.

During the literature review, we identified several repositories that provide collections of spam emails. However, since our study specifically targets recent efforts and datasets that focus exclusively on phishing emails, these general spam datasets were not suitable and were therefore not used. Nonetheless, these datasets may still be valuable for researchers working on broader spam detection tasks. Wang et al. \cite{wang2013study} collected a repository of 5.1 million spam emails collected over 15 years (1998-2013). Their work included statistical analysis of various email header fields, such as content type, message length, embedded URLs, and HTML attachments. The authors have continuously updated the repository, and its latest version includes a significantly larger set of spam samples.

The Apache SpamAssassin dataset \cite{spamassassin_corpus} provides a repository for spam email classification, divided into three categories: spam, easy ham, and hard ham. The easy ham set contains legitimate emails that are clearly distinguishable from spam, as they lack typical spam characteristics. In contrast, the hard ham set includes legitimate emails that exhibit features commonly found in spam, such as HTML formatting, colored text, and atypical markup. The dataset includes a total of 4,150 ham emails, 3,900 labeled as easy ham and 250 as hard ham, along with 1,897 spam emails. Oard et al. \cite{oard2015avocado} compiled the Avocado Research Email Collection, comprising approximately 1.2 million legitimate emails and attachments from 279 user accounts of a defunct information technology company. These accounts primarily belong to employees, with some representing shared or system accounts. The data were extracted from Microsoft Outlook Personal Storage Table (PST) files using the libpst tool and include both structured metadata and message content. This dataset is widely utilized for social network analysis, email classification, and e-discovery research.

In summary, our study identifies three key gaps in the current literature on phishing email detection, each of which presents important challenges for the development of efficient and robust detection systems:

\begin{itemize}
\item While many existing datasets offer a wide range of traditional phishing and legitimate samples, models trained exclusively on these datasets fail to detect sophisticated phishing emails generated by LLMs. These AI-generated emails exhibit high linguistic fluency and closely mimic legitimate communication, making them harder to detect.
\item Models trained solely on LLM-generated samples overlook important structural and linguistic patterns characteristic of traditional phishing attacks. To address this, our study proposes a hybrid dataset that integrates both human-crafted and LLM-generated phishing emails, and offers a more realistic and diverse training corpus.
\item While transformer-based architectures such as BERT, TinyBERT, and MobileBERT, including prompt-based variants like PSC-BERT \cite{gui2024psc}, achieve strong detection performance due to their large-scale pretraining on massive corpora, their computational complexity limits real-world deployment. Our approach tackles this issue by distilling knowledge from a fine-tuned MobileBERT into a lightweight BiLSTM model, thereby improving semantic generalization while enabling efficient deployment in resource-constrained environments. Related work also explores transformer-CNN hybrids for phishing detection at the URL level, such as the TCURL model proposed by Wang et al. \cite{wang2022tcurl}, which further demonstrates the adaptability of transformer-based techniques across phishing modalities.

\end{itemize}

%% file: Data_agmentation.tex
\label{dataset_generation}

\begin{figure}[!h]
    \centering
    \includegraphics[width=0.45\textwidth]{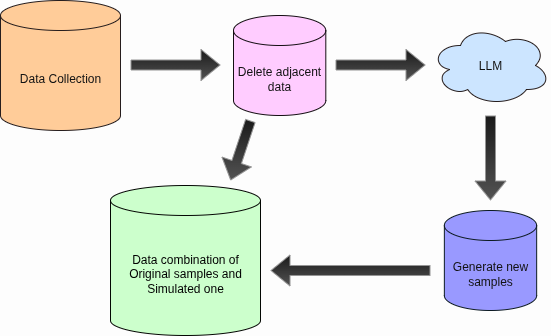}
    \caption{\small The process of creating a combined dataset for training a phsihing email detection.}
    \label{fig:original_and_simulated_phishing}
\end{figure}

To develop robust and generalizable phishing detection models, it is essential to train on a diverse and representative dataset that captures both traditional and next-generation phishing techniques \cite{nahmias2024prompted}. This section presents our methodology for constructing such a dataset by integrating real-world email samples with synthetically generated phishing content produced by Large Language Models (LLMs). Figure~\ref{fig:original_and_simulated_phishing} It shows the entire process of preparing the dataset. We first collect and preprocess publicly available datasets containing legitimate and phishing emails. These datasets are then used to guide the generation of synthetic phishing samples that are contextually relevant and grammatically coherent. The following subsections detail the preprocessing of real email data, the pipeline for LLM-based synthetic generation, and the integration of auxiliary datasets to form a comprehensive and realistic phishing corpus.

\subsection{Basic Datasets}

The datasets presented in this section are either publicly accessible for research purposes or obtained through formal access requests supported by institutional documentation. All datasets are used in compliance with their respective licensing terms and are exclusively employed for academic research on phishing email detection. A summary of these datasets is presented in Table \ref{phishing_datasets}, and further details for each dataset are provided below:

\begin{itemize}

\item  \textbf{Cambridge Phishing Dataset} \cite{CambridgeCybercrimeCentre}: Published by the Cambridge Cybercrime Centre at the University of Cambridge, this dataset includes an extensive collection of spam and phishing emails. It reflects significant annual variation, with recent updates exceeding 2,000 emails per month. For our study, we utilized the latest release from December 2024, which includes raw email content along with associated metadata.

\item  \textbf{Nazario Phishing Corpus \cite{gonzalez2011phishing}}: This dataset comprises 4,555 phishing emails collected over several years and released publicly in 2005. It has been widely used in academic research as a benchmark for evaluating phishing detection methods. The emails originate from diverse sources and represent various phishing tactics, including credential harvesting, embedded malicious links, and socially engineered content.

\item  \textbf{Chakraborty Dataset} \cite{subhajournal_phishingemails}: This dataset includes approximately 1,800 email samples, with 61\% labeled as phishing and 39\% as legitimate. It provides only the email body text, without metadata such as headers or timestamps. While limited in size and scope, this dataset serves as a baseline benchmark for evaluating text-based phishing detection models.

\item  \textbf{Phishing Pot Dataset} \cite{phishing_pot}: This dataset contains 4,352 real phishing emails captured via a honeypot infrastructure specifically designed to attract phishing attempts. Each email is provided along with metadata, including headers, timestamps, and full message content. 

\item  \textbf{Enron Dataset} \cite{EnronEmailDataset}: Collected by the CALO Project (A Cognitive Assistant that Learns and Organizes), this dataset comprises approximately 500,000 legitimate emails from 150 Enron employees, primarily senior executives. It is one of the most widely used corpora for email research and includes complete message content along with metadata such as sender, recipient, and timestamps.

\end{itemize}

\begin{table*}[h]
    \centering
    \caption{Datasets Used}
    \renewcommand{\arraystretch}{1.3}
    \resizebox{\textwidth}{!}{%
    \begin{tabular}{
        >{\centering\arraybackslash}p{0.35cm}  
        >{\centering\arraybackslash}p{3.5cm}
        >{\centering\arraybackslash}p{1.5cm}
        >{\centering\arraybackslash}p{1.5cm}
        >{\centering\arraybackslash}p{0.9cm}
        >{\centering\arraybackslash}p{0.7cm}  
        >{\centering\arraybackslash}p{5cm}
    }
        \hline
        \rowcolor[HTML]{D3D3D3}
        \textbf{\scriptsize No} & 
        \textbf{\scriptsize Dataset Name} & 
        \textbf{\scriptsize Phishing Emails} & 
        \textbf{\scriptsize Legitimate Emails} & 
        \textbf{\scriptsize Total} & 
        \textbf{\scriptsize Year} & 
        \textbf{\scriptsize Metadata Available} \\
        \hline

        \rowcolor[HTML]{F0F0F0} \scriptsize 1 & \scriptsize Cambridge Phishing & \scriptsize 78154 & - & \scriptsize 78154 & \scriptsize 2024 & \tiny Full headers: From, To, Subject, Date, Message-ID, Received, Return-path, CCC tags (HASH, DATE), MIME info  \\

        \rowcolor[HTML]{FFFFFF} \scriptsize 2 & \scriptsize Nazario Phishing Corpus & \scriptsize 4487 & - & \scriptsize 4487 & \scriptsize 2005 & \tiny Full headers (From, To, Subject, Received, Auth Results, Spam Score, X-PHP-Script), MIME info  \\

        \rowcolor[HTML]{F0F0F0} \scriptsize 3 & \scriptsize Chakraborty Phishing & \scriptsize 7323 & \scriptsize 11283 & \scriptsize 18606 & \scriptsize 2023 & \tiny Body text only \\

        \rowcolor[HTML]{FFFFFF} \scriptsize 4 & \scriptsize Phishing Pot & \scriptsize 4351 & - & \scriptsize 4351 & \scriptsize 2024 & \tiny From, To, Subject, Date, Message-ID, Reply-To, Return-Path, Content-Type, SPF, SCL, Priority, Received, X-MS headers \\

        \rowcolor[HTML]{F0F0F0} \scriptsize 5 & \scriptsize Enron Dataset & - & \scriptsize 517,402 & \scriptsize 517,402 & \scriptsize 2004 & \tiny MsgID, Date, From, To, Subj, MimeVer, C-Type, C-Encoding, X-From, X-To, X-Fold, X-Orig, X-FName \\

        \hline
        \rowcolor[HTML]{FFFFFF} \scriptsize 6 & \scriptsize Combined Real-World Samples & \scriptsize 94,315 &  \scriptsize 528,685 & \scriptsize 623,000 & - & \tiny Varies \\

        \rowcolor[HTML]{E6F7FF} \scriptsize 7 & \scriptsize LLM-Generated Samples & \scriptsize 41333 & \scriptsize 6242 & \scriptsize 47575 & \scriptsize 2025 & \tiny Full text only \\
        \hline
    \end{tabular}
    }
    \label{phishing_datasets}
\end{table*}

\subsection{Generated Synthetic Samples by LLM}

After the collection of a diverse corpus of legitimate and phishing emails, this section describes the process for generating synthetic phishing samples using closed-source commercial LLM services such as OpenAI and DeepSeek. The goal is to produce high-quality adversarial examples that reflect a wide range of linguistic styles, attack strategies, and semantic structures commonly observed in phishing emails \cite{das2019automated}. These LLM-generated samples are designed to bypass traditional detection systems that depend on static rules or manually engineered features. Their fluent language, contextual coherence, and adaptive phrasing significantly increase the likelihood of user deception, posing a greater challenge for conventional phishing detection systems \cite{guo2024x}.

The assessment of LLM-generated phishing emails' semantic proximity to legitimate content involved calculating cosine similarity scores between each legitimate email and both original and simulated phishing messages. The TF-IDF vector representations were built from a vocabulary that included all emails and cosine similarity was calculated between each legitimate email and the phishing corpora. The resulting scores represent the average semantic closeness for each pairwise comparison.

\begin{figure}[H]
\centering
\includegraphics[width=0.9\linewidth]{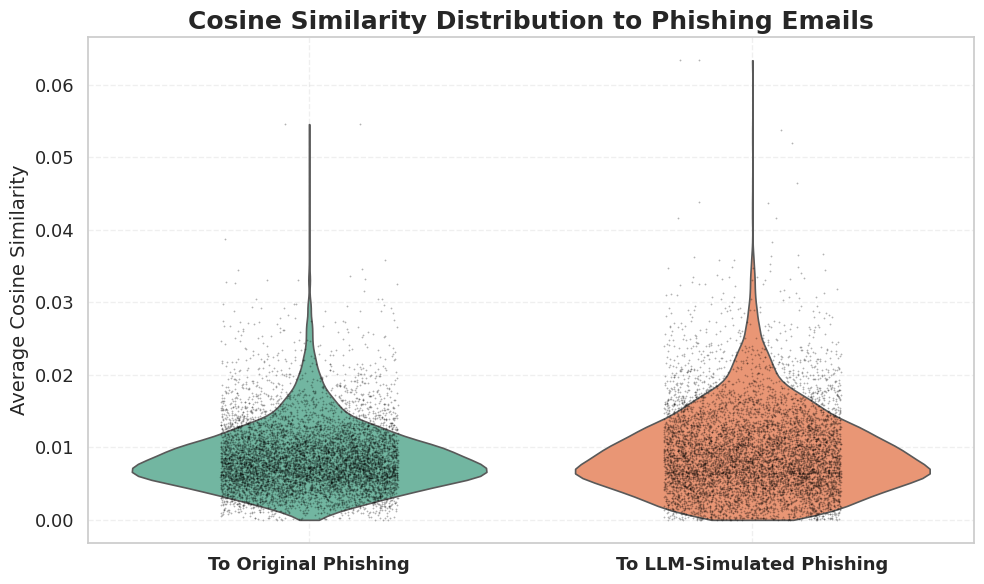}
\caption{The analysis examines the cosine similarity between actual emails and phishing examples. The violin plot displays the distribution of semantic similarity scores which were calculated through TF-IDF and cosine similarity. The semantic similarity between LLM-generated phishing emails (orange) and legitimate emails exceeds that of traditional phishing samples (green) which makes them more likely to evade detection.}
\label{fig:similarity_distribution}
\end{figure}

The distribution of similarity scores appears in Figure ~\ref{fig:similarity_distribution} through a violin plot which includes individual sample points. The similarity scores of LLM-generated phishing emails show a rightward shift compared to original phishing messages which indicates they share more lexical and contextual elements with legitimate emails. The high degree of similarity in LLM-generated content indicates its advanced nature and its ability to evade traditional detection systems. The results demonstrate the necessity for detection systems to use semantically enriched and context-aware models to identify these sophisticated deceptive threats.

\subsubsection{Phishing-Level Scoring for Validation and Controlled Text Generation}

To validate and refine the quality of generated phishing samples, we introduced a question-driven scoring mechanism that leverages the reasoning capabilities of LLMs for classification. This approach formulates a set of targeted questions, each aligned with key phishing indicators such as urgency, suspicious hyperlinks, authority impersonation, and requests for confidential information. These questions are designed to evaluate specific linguistic or contextual traits typically observed in phishing attacks, enabling controlled adjustment of adversarial features in generated emails.

For each generated email, we submitted a series of binary and scaled prompts to the LLM to evaluate specific phishing-related characteristics. These prompts were designed to elicit judgments on features commonly associated with phishing. Examples include:

\begin{itemize}
\item Does this email convey a sense of urgency or panic?
\item How much flattery is present in this email?
\item How suspicious is the embedded link?
\item How closely does this email resemble a marketing message?
\item Does the email address the recipient by name or include overly specific details?
\item To what extent does the message imply consequences for inaction (authority)?
\item Does the email request account updates or signature actions via a link?
\item How strongly does the email pressure the recipient to click a link urgently?
\end{itemize}

\begin{figure*}[!h]
    \centering
    \subfloat[]{\includegraphics[width=8cm]{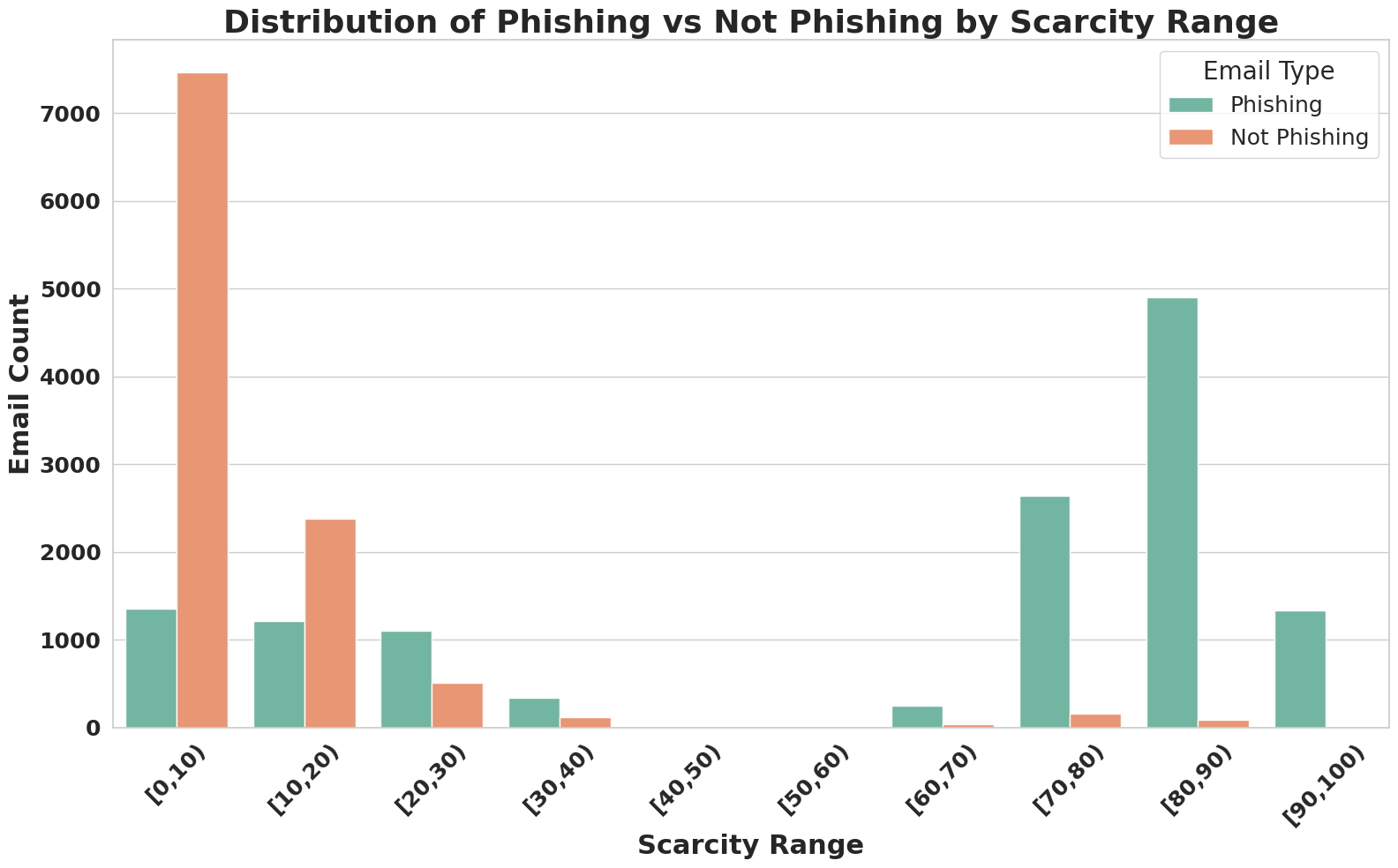} }
        \subfloat[]{{\includegraphics[width=8cm]{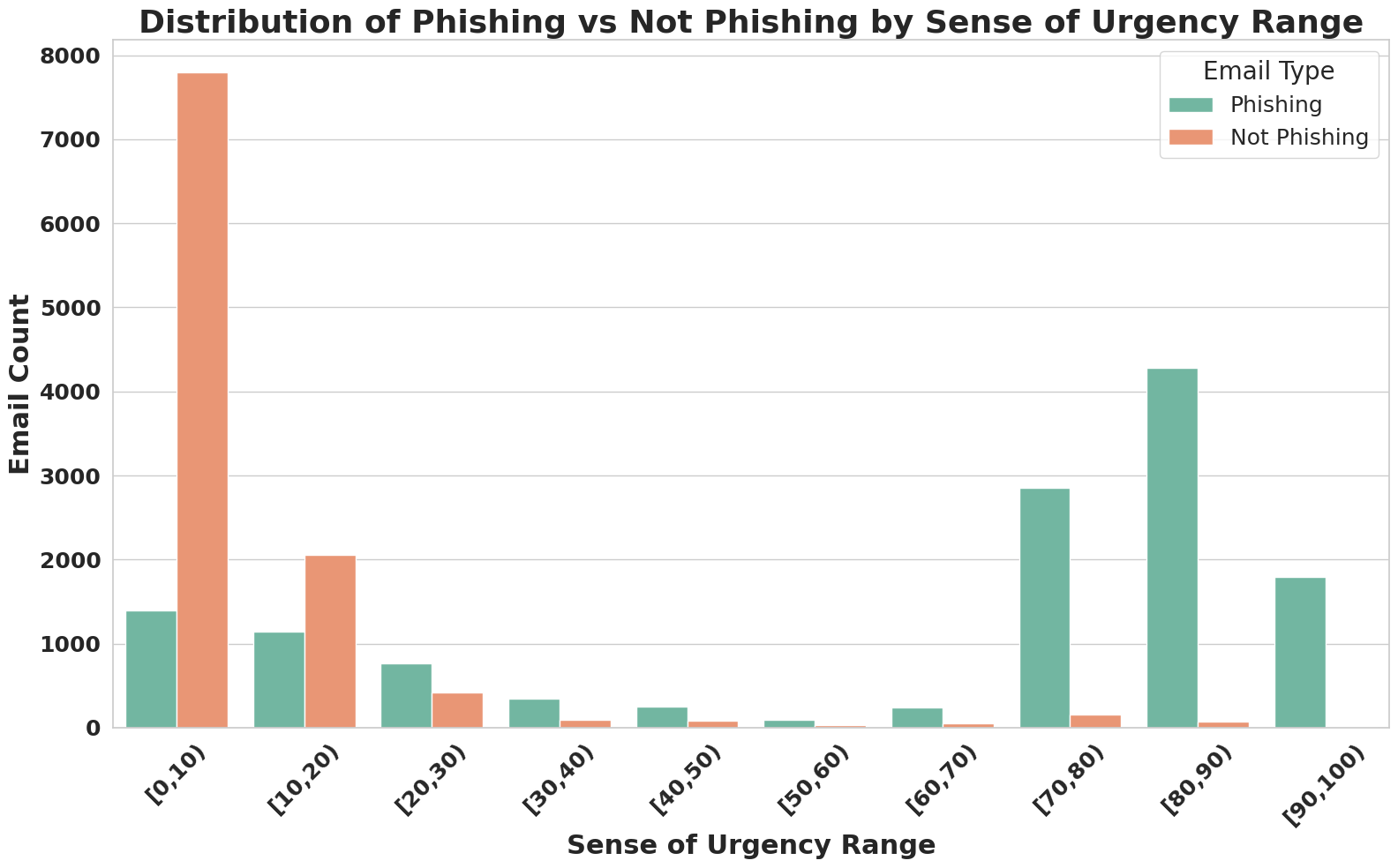} }}
    \caption{\small LLM-based evaluation of phishing severity: (a) scoring based on scarcity cues, and (b) scoring based on urgency cues.}
    \label{SuspiciousLink_vs_SenseOfUrgency}
\end{figure*}

The LLM’s responses were treated as soft classification signals, which contribute to confidence-weighted scores for phishing likelihood. The discriminative power of each prompt was assessed based on its alignment with ground truth labels. As illustrated in Figure \ref{SuspiciousLink_vs_SenseOfUrgency}, certain prompts, such as those assessing link suspiciousness or urgency, demonstrated strong alignment with phishing indicators, thereby exhibiting high predictive utility. In contrast, other prompts showed weaker correlation, underscoring the need for careful prompt selection that balances linguistic structure with semantic relevance to phishing behavior.

To quantify phishing intent, we aggregated the LLM’s responses across all prompts to compute a composite phishing-level score for each email. This score was used for both sample validation (filtering low-confidence synthetic emails) and controlled generation (guiding the model to produce samples with specific phishing attributes). Moreover, the multi-prompt evaluation provided interpretability by identifying specific tactics present in each message, such as urgency, impersonation, or credential harvesting. This framework demonstrates that LLMs not only generate adversarial phishing content, but can also perform self-assessment via prompt-guided reasoning, offering a novel feedback loop to improve both sample quality and training effectiveness.

\subsubsection{Sample Selection and Deduplication} To ensure a fair and diverse evaluation, we initially generated a balanced set of legitimate and phishing samples from the 25,000 collected emails across multiple sources. To reduce semantic redundancy and avoid evaluation bias, we conducted a similarity analysis to identify overlapping or highly similar samples. Approximately 10,000 duplicates were removed through this process, resulting in a final corpus of 14,000 distinct emails. This deduplication step ensures that the dataset preserves linguistic diversity and more accurately reflects the variability observed in real-world phishing scenarios.

%% file: methodology.tex
\label{Methodology}

To develop a context-aware baseline for phishing detection, we implemented and evaluated multiple configurations of Long Short-Term Memory (LSTM) models across  three progressive enhancement stages, after completing comprehensive preprocessing and tokenization of the email data. First, we establish a baseline model by integrating BiLSTM with a custom single-head attention mechanism to emphasize key linguistic patterns relevant to phishing. Next, we extend this architecture with a multi-head attention module, enabling the model to simultaneously capture multiple semantic dependencies across the input sequence. Finally, knowledge distillation was applied by transferring representational knowledge from a fine-tuned MobileBERT model into the multi-head attention BiLSTM. This multi-stage design leverages the sequential modeling capabilities of recurrent networks alongside the expressiveness of transformer-based attention mechanisms, by achieving high detection performance while substantially reducing computational overhead compared to even the most lightweight Large Language Models (LLM) variants.


\begin{figure*}[!h]
\begin{center}
\centering
{\includegraphics[width=6 in]{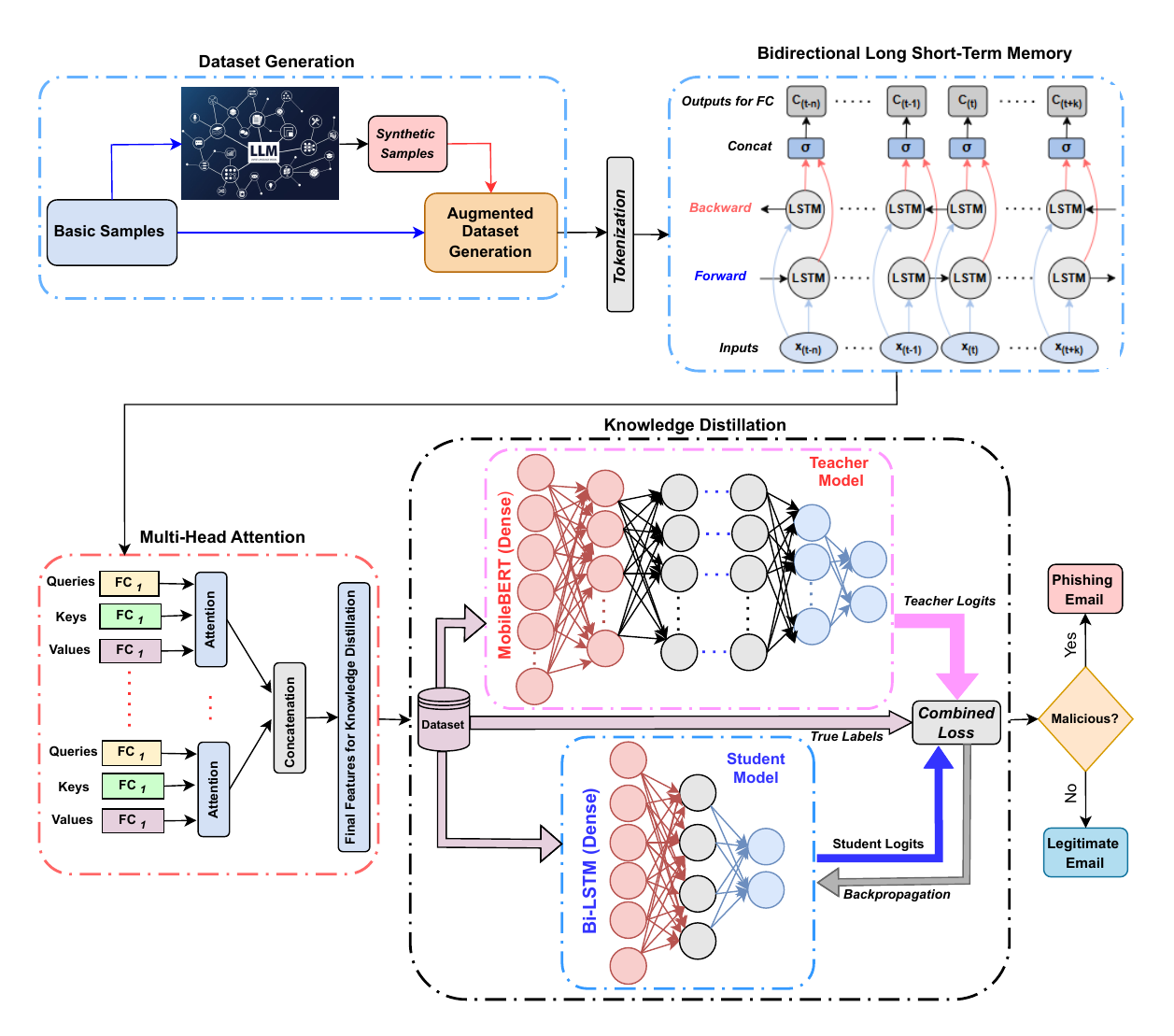}}\\
\vspace{-0.2cm}
\caption{Overview of the proposed LLM-aware phishing email detection architecture.}
\label{framework}
\end{center}
\end{figure*}

Figure \ref{framework} illustrates the overall architecture of the proposed phishing email detection framework. The system begins with a hybrid dataset comprising real and LLM-generated phishing emails. Emails are preprocessed and tokenized before being fed into a BiLSTM model enhanced with multi-head attention. Knowledge distillation is applied from a fine-tuned MobileBERT teacher model to improve the student model’s generalization while maintaining efficiency for deployment.


\subsection{Tokenization Process}

To capture contextual and semantic relationships within email text, we utilize the Word2Vec embedding model. Raw emails are preprocessed through steps such as lowercasing, digit removal, and stopword filtering to produce cleaned token sequences. These sequences are then fed into Word2Vec, which generates 100-dimensional dense vector representations by optimizing either the Skip-Gram or Continuous Bag-of-Words (CBOW) \cite{mikolov2013efficient} objective to approximate token co-occurrence statistics across the combined corpus. Each email is subsequently encoded as the element-wise mean of its token vectors, resulting in a fixed-length feature vector that preserves overall semantic meaning.

\subsection{Long Short-Term Memory (LSTM)}

The sequential data processing capabilities of LSTM networks make them highly effective because they can detect long-range dependencies. The memory function of LSTM networks proves essential for phishing detection because important cues such as urgency and impersonation and suspicious links can appear throughout an email. The first stage of our model uses LSTM cells to analyze contextual flow before we move on to bidirectional enhancement and attention mechanisms in the following stages.

The LSTM cell controls information flow through gating mechanisms in each of its components. The \textit{forget gate} functions to eliminate unimportant input information from memory so the model can concentrate on meaningful semantic tokens. It is computed as:

\begin{equation}
f_{t} = \sigma(W_{f} \cdot [h_{t-1}, x_{t}] + b_f)
\end{equation}

where $x_t$ is the current input, $h_{t-1}$ is the previous hidden state, and $\sigma$ is the sigmoid activation function.

The \textit{input gate} determines which new information to store through the candidate cell state $\hat{C}_t$.

\begin{equation}
i_{t} = \sigma(W_{i} \cdot [h_{t-1}, x_{t}] + b_i) 
\end{equation}
\begin{equation}
\hat{C}_t = \tanh(W_C \cdot [h_{t-1}, x_t] + b_C)
\end{equation}

The filter function \( i_t \) selects which candidate information \( \hat{C}_t \) will be stored in the cell's memory. The model requires this function to learn new deceptive patterns in phishing detection while discarding unimportant tokens.

The internal memory $C_t$ is then updated by combining retained past memory with relevant new content:

\begin{equation}
C_t = f_t \odot C_{t-1} + i_t \odot \hat{C}_t
\end{equation}

where $\odot$ denotes element-wise multiplication.

The \textit{output gate} produces the hidden state $h_t$ which moves to the following LSTM cell or classifier:

\begin{equation}
o_t = \sigma(W_o \cdot [h_{t-1}, x_t] + b_o), \quad
h_t = o_t \odot \tanh(C_t)
\end{equation}

\begin{figure}[!h]
\centering
\includegraphics[width=3.3in]{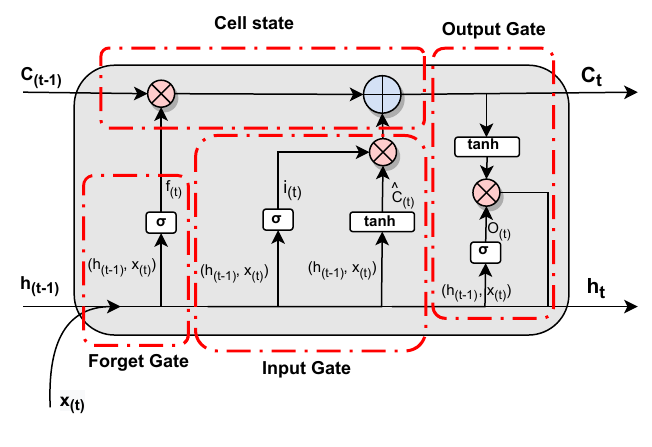}
\vspace{-0.2cm}
\caption{The internal structure of an LSTM cell.}
\label{LSTMcell}
\end{figure}

The output system requires knowledge about sequential and contextual understanding to detect deceptive cues in phishing emails. The internal architecture of the system is shown in Figure ~\ref{LSTMcell}.

\subsection{Bidirectional LSTM (BiLSTM)}
\begin{figure}[!h]
\begin{center}
\centering
{\includegraphics[width= 3.4 in]{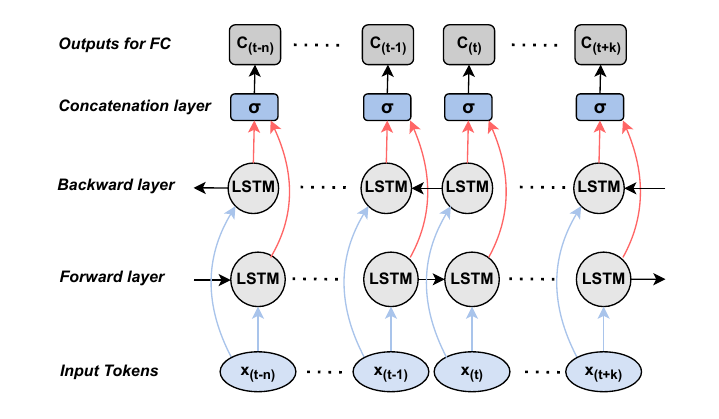}}\\
\caption{The architecture of Bi-directional LSTM.}
\label{Bi-LSTM-unit}
\end{center}
\end{figure}

Traditional LSTM networks process input sequences in a single direction (typically left to right), which restricts the model to using only the past context (i.e., from $x_1$ to $x_{t-1}$) when making predictions at time step $t$. However, in phishing email detection, the interpretation of a sentence often depends on both its preceding and following words. This unidirectional approach can thus lead to the loss of valuable semantic information. To address this limitation, we employ a Bidirectional LSTM architecture, which processes the input sequence in both forward and backward directions. This design allows the model to effectively capture dependencies from both the past and future, enabling a more comprehensive understanding of the sentence structure and intent. Additionally, the bidirectional structure enhances the model’s ability to detect reverse dependencies, understand full sentence context, and generalize more effectively across traditional and LLM-generated phishing samples. Figure \ref{Bi-LSTM-unit} illustrates the internal structure of the BiLSTM unit, showing how information is propagated and combined from both directions to form context-aware representations.

Given an input sequence $X = {x_1, x_2, \dots, x_n}$, the BiLSTM independently computes forward and backward hidden states at each time step $t$ as follows:

\begin{equation}
\overrightarrow{h}_t = \text{LSTM}_f(x_t, \overrightarrow{h}_{t-1})
\label{eq:lstm_forward}
\end{equation}

\begin{equation}
\overleftarrow{h}_t = \text{LSTM}_b(x_t, \overleftarrow{h}_{t+1})
\label{eq:lstm_backward}
\end{equation}

The final hidden state at time step \( t \) is obtained by concatenating the forward and backward hidden states:

\begin{equation}
h_t = [\overrightarrow{h}_t \parallel \overleftarrow{h}_t]
\label{eq:bidirectional_concat}
\end{equation}

where $\parallel$ denotes vector concatenation. This bidirectional structure allows the model to incorporate both preceding and succeeding contextual information.

\subsection{Attention-Enhanced Recurrent Architecture (Single head)}

Although the BiLSTM architecture captures bidirectional contextual dependencies, it treats all tokens equally when generating representations for classification. In phishing detection, however, specific phrases, such as \textit{verify your account}, \textit{limited time offer}, or \textit{Urgent: Your account will be suspended}, carry significantly higher semantic weight in signaling malicious intent. To better emphasize these critical regions, we integrate a single-head attention mechanism on top of the BiLSTM outputs. This mechanism assigns attention weights to individual tokens based on their relevance, enabling the model to prioritize more informative words while down-weighting irrelevant or benign content. As a result, the model produces a context-aware sentence-level representation that encapsulates the most phishing-relevant information, and improve both classification performance and interpretability.

The attention mechanism operates by projecting the BiLSTM output sequence into three learned subspaces: Query ($Q$), Key ($K$), and Value ($V$). These are computed through trainable linear transformations applied to the BiLSTM outputs:

\begin{equation}
Q = W_q \cdot h, \quad K = W_k \cdot h, \quad V = W_v \cdot h
\label{eq:qkv}
\end{equation}

where $h$ denotes the hidden states produced by the BiLSTM, and $W_q$, $W_k$, and $W_v$ are learnable weight matrices. To compute the attention scores, we employ the scaled dot-product attention mechanism:

\begin{equation}
\text{Attention}(Q, K, V) = \text{softmax}\left(\frac{QK^\top}{\sqrt{d_k}}\right)V
\label{eq:scaled_dot_product}
\end{equation}

Here, $d_k$ represents the dimensionality of the key vectors. The scaling factor $\sqrt{d_k}$ prevents the softmax function from producing extremely small gradients, which can impair learning. This mechanism enables the model to dynamically weight token representations based on their relevance, allowing it to selectively amplify key phrases commonly found in phishing emails while suppressing irrelevant information.

\subsection{Classification Layer}

The output of the attention mechanism is passed to a fully connected (Dense) layer with a sigmoid activation function. This layer maps the aggregated features to a scalar value $\hat{y}_i \in (0, 1)$, which represents the probability that the input email is phishing. The formulation is:

\begin{equation}
\hat{y}_i = \sigma(W_o \cdot \text{Attention}(Q, K, V) + b_o)
\end{equation}

where $W_o$ and $b_o$ are the learnable weight and bias parameters of the output layer, and $\sigma$ denotes the sigmoid activation function. To train the model, we minimize the binary cross-entropy loss between the predicted value $\hat{y}_i$ and the true label $y_i$:

\begin{equation}
L(\theta) = -\frac{1}{N} \sum_{i=1}^{N} \left[ y_i \cdot \log(\hat{y}_i) + (1-y_i) \cdot \log(1-\hat{y}_i) \right]
\end{equation}

The model is optimized using the Adam optimizer with an initial learning rate of 0.001. Dropout is applied before the output layer to reduce overfitting and enhance generalization. Table \ref{BiLSTM_with_attention} summarizes the architectural components of the proposed model.

\begin{table}[!h]
\caption{The internal components of the proposed Context-aware BiLSTM with Multi-Head Attention model}
\centering
\label{BiLSTM_with_attention}
\begin{tabular}{p{0.23\linewidth} p{0.18\linewidth} p{0.43\linewidth}} 
  \bottomrule
  \textbf{\tiny Layers} & \textbf{\tiny Output shape} & \textbf{\tiny Description}  \\
  \bottomrule
  \tiny Input layer & \tiny (Batch size, $T$, $F$) & \tiny Sequence of input tokens or feature embeddings; $T$ is sequence length, $F$ is feature dimension. \\
    \tiny Bidirectional LSTM & \tiny (Batch size,128) & \tiny BiLSTM with  128 units per direction, activation = tanh. \\
    \tiny Custom Multi-Head Attention & \tiny (Batch size, $T$, $H\times D$) & \tiny Multi-head attention with $H=4$ heads and $D=64$ key/query/value dims; Q/K/V projections to $H\times D=256$, followed by output projection. \\
    \tiny Dropout & \tiny (Batch size, 128) & \tiny Dropout rate = 0.5 applied to the BiLSTM output. \\
  \tiny Fully-connected (Dense) & \tiny (Batch size, 1) & \tiny Single sigmoid-activated unit for binary classification. \\ 
  
  \tiny  Loss function & \tiny - & \tiny  Binary cross-entropy used for optimization objective. \\
  \tiny  Optimizer & \tiny - & \tiny  Adam optimizer with initial learning rate of 0.001.\\

  \tiny Batch Size and Epochs & \tiny - & \tiny  Batch Size = 32, Epochs = 5 \\
  
  \bottomrule
\end{tabular}
\end{table}

\subsection{BiLSTM Enhanced with Multi-Head Attention}

While single-head attention improves the semantic representation of BiLSTM outputs by emphasizing key tokens, a robust phishing detection requires a more expressive mechanism to capture subtle and distributed pattern indicators. Modern phishing emails are crafted to closely mimic legitimate correspondence, embedding nuanced indicators, such as emotional manipulation, spoofed branding, and urgency-inducing language, across various parts of the email. Relying on a single attention vector limits the model’s ability to capture such diverse signals. To address this, we incorporate a multi-head attention mechanism to enable the model to attend to multiple phishing-relevant patterns in parallel. Each attention head independently learns to focus on different semantic aspects of the input.

Technically, multi-head attention model can capture complex phrase-level dependencies that single-head attention may overlook. For instance, in the sentence \textit{Please confirm your password to avoid account suspension}, one head might focus on \textit{confirm your password}, while another emphasizes \textit{avoid account suspension}. Although semantically different, both phrases are strongly indicative of phishing and benefit from being simultaneously represented. By combining these multiple perspectives, the model gains a richer, more comprehensive understanding of phishing intent.

For a given input sequence X, the multi-head attention mechanism computes multiple parallel attention distributions, each focusing on different representational subspaces. The outputs of these heads are concatenated:

{\small
\begin{equation}
\text{MultiHead}(X) = \text{Concat}(\text{head}_1, \text{head}_2, \dots, \text{head}_n) 
\label{eq:multihead_attention}
\end{equation}
}

Each individual attention head is computed using the scaled dot-product attention mechanism, applied to linearly projected queries, keys, and values:

\begin{equation}
\text{head}_i = \text{Attention}(XW^Q_i, XW^K_i, XW^V_i)
\label{eq:head_i}
\end{equation}

Here, $W^Q_i$, $W^K_i$, and $W^V_i$ represent the learned projection matrices for the $i^{th}$ head's queries, keys, and values, respectively. After concatenation, the multi-head output passes through a shared linear transformation to produce a unified representation:

\begin{equation}
\text{Output} = \text{MultiHead}(X) \cdot W^O + b^O
\label{eq:final_projection}
\end{equation}

This final embedding integrates multiple semantic perspectives from different heads and serves as input to a dense classification layer.  The model computes the prediction $\hat{y}_i \in (0,1)$ using a sigmoid activation function:

\begin{equation}
\hat{y}_i = \sigma(W^{\text{dense}} \cdot \text{Output} + b^{\text{dense}})
\label{eq:final_dense_classification}
\end{equation}

Here, $W^O$ and $b^O$ denote the projection weights and bias of the multi-head attention output, while $W^{\text{dense}}$ and $b^{\text{dense}}$ define the parameters of the final classification layer. The remainder of the classification pipeline, including the use of sigmoid activation and binary cross-entropy loss, follows the same methodology as described earlier for the single-head attention model.


\subsection{MobileBERT Architecture (Teacher Model)}

In this section, we establish a strong baseline using a transformer-based architecture. MobileBERT is adopted as the teacher model to assess its effectiveness in phishing email classification. This foundation enables a reliable comparison for subsequent steps involving feature refinement, attention mechanisms, and model distillation into more lightweight architectures.

\subsubsection{Baseline Architecture} 

The baseline model is based on MobileBERT \cite{sun2020mobilebert}, a compact and efficient variant of the BERT architecture designed for deployment in resource-constrained environments. MobileBERT maintains the core transformer-based structure of BERT while optimizing for memory and computational efficiency. Given an input sequence $X = \{ x_1, x_2, \dots, x_n \}$, each token $x_i$ is first mapped to a dense vector $e_i$ using the embedding layer:

\begin{equation}
e_i = \text{Embed}(x_i)
\label{eq:embedding}
\end{equation}

The embedded token sequence is then passed through $L$ transformer layers. Each layer incorporates multi-head self-attention and position-wise feed-forward sublayers. The transformation at the $l$-th layer is expressed as:

\begin{equation}
h^l = \text{TransformerLayer}(h^{(l-1)})
\label{eq:transformer}
\end{equation}

where $h^l$ denotes the hidden state at layer $l$, and $h^0$ corresponds to the initial token embeddings $e$.


\subsubsection{Data Processing Pipeline}
The preprocessing pipeline comprises a sequence of carefully structured steps to ensure the model receives high-quality input. The process begins with text normalization, where each email undergoes cleaning and formatting to eliminate noise such as HTML tags, special characters, and extraneous whitespace. 

Following normalization, the text is tokenized using the WordPiece algorithm, which segments the input into subword units suitable for transformer-based models. For an input sequence of length $n$, the tokenization process yields:

\begin{equation}
    T = \{ t_1, t_2, \dots, t_m \}
    \label{eq:T_set}
\end{equation}

where $m \leq 512$, conforming to the maximum input length constraint of the MobileBERT architecture. Special tokens \texttt{[CLS]} and \texttt{[SEP]} are appended at the beginning and end of the sequence, respectively. These serve as classification anchors and delimit sentence boundaries for downstream processing.

\subsubsection{Training Objective and Optimization}
The model parameters, denoted by $\theta$, are optimized using the Adam optimizer with a variable learning rate schedule and an initial warmup phase \cite{sun2020mobilebert}. The objective is to minimize the binary cross-entropy loss, defined as:

{\small
\begin{equation}
    L(\theta) = -\frac{1}{N} \sum_{i=1}^{N} \left[ y_i \log p_{\theta}(y_i | X_i) + (1 - y_i) \log (1 - p_{\theta}(y_i | X_i)) \right]
    \label{eq:loss_function}
\end{equation}
}

Here, $N$ represents the batch size, $y_i$ is the true binary label for the $i^\text{th}$ email, and $p_{\theta}(y_i \mid X_i)$ denotes the model’s predicted probability. This formulation ensures effective gradient updates during fine-tuning on phishing detection tasks.




\subsection{Knowledge Distillation from MobileBERT into Multi-Head Attention BiLSTM}

Although the BiLSTM architecture augmented with multi-head attention increases the model's capacity to capture contextual and semantic dependencies, it remains limited by the size and diversity of the available training data. Even after aggregating multiple phishing datasets from various sources, the overall volume and linguistic variability are still considerably smaller than those found in the large-scale corpora used to pretrain LLMs. However, the deployment of LLMs requires specialized resources and is often infeasible in resource-constrained environments. Even though a much lighter version of BERT, such as MobileBERT, was selected for training and detection, its training time and resource demands make it impractical for phishing detection in real-world settings, despite achieving higher accuracy compared to lightweight BiLSTM models.

The limitations of MobileBERT for real-world deployment, despite its strong detection performance, motivated us to bridge this gap by further enhancing our latest BiLSTM model, augmented with multi-head attention, through knowledge distillation from the fine-tuned MobileBERT model. This technique transfers representational knowledge from the fine-tuned MobileBERT model into the BiLSTM-based architecture. As a result, the distilled BiLSTM model, retain MobileBERT’s semantic understanding and generalization capabilities while offering improved efficiency for use in resource-constrained environments.

Knowledge distillation, first introduced by Hinton et al. \cite{hinton2015distilling}, enables the transfer of knowledge from a complex \textit{teacher} model to a more compact \textit{student} model. In addition to learning from ground-truth labels, the student is trained to replicate the soft probability distributions produced by the teacher. As illustrated in Figure \ref{framework}, the proposed method distills knowledge from a fine-tuned MobileBERT model into a BiLSTM model augmented with multi-head attention.

The overall distillation process is guided by two learning objectives, combined as follows:

\begin{equation}
L_{\text{distill}} = \alpha \cdot L_{\text{hard}} + (1 - \alpha) \cdot \tau^2 \cdot L_{\text{soft}}
\label{eq:distillation_loss}
\end{equation}

Here, $L_{\text{hard}}$ denotes the standard cross-entropy loss computed with ground-truth labels, while $L_{\text{soft}}$ refers to the Kullback-Leibler divergence between the softened output distributions of the teacher and student models. The parameter $\alpha \in [0, 1]$ balances the contributions of the two objectives. The temperature parameter $\tau$ is used to soften the logits, making the output distributions more informative for the student model during training.

The specific loss functions are defined as follows. The hard loss corresponds to the cross-entropy between the ground truth labels and the student's predicted probabilities:

\begin{equation}
L_{\text{hard}} = -\sum_i y_i \cdot \log(p_i^S)
\label{eq:hard_loss}
\end{equation}

The soft loss is the Kullback-Leibler (KL) divergence \cite{kim2021comparing} between the softened output distributions of the teacher and student models:

\begin{equation}
L_{\text{soft}} = \text{KL}\left(\sigma\left(\frac{z^T}{\tau}\right) \bigg\| \sigma\left(\frac{z^S}{\tau}\right)\right)
\label{eq:soft_loss}
\end{equation}

This expression can be expanded as:

\begin{equation}
L_{\text{soft}} = \sum_i \sigma\left(\frac{z_i^T}{\tau}\right) \cdot \log\left(\frac{\sigma\left(\frac{z_i^T}{\tau}\right)}{\sigma\left(\frac{z_i^S}{\tau}\right)}\right)
\label{eq:kl_divergence_expanded}
\end{equation}

Here, $y_i$ denotes the true label, $p_i^S$ is the predicted probability from the student model, and $z^T$ and $z^S$ represent the logits from the teacher and student models, respectively. The function $\sigma$ is the softmax activation, defined as $\sigma(z_i) = \frac{e^{z_i}}{\sum_j e^{z_j}}$.

We balance the influence of hard and soft targets by setting $\alpha = 0.5$ and $\tau = 2.0$ in our implementation, which ensures sufficient information flow through the softened output distributions.


\subsubsection{Embedding Transfer and Student Architecture}
To facilitate efficient knowledge transfer, the embedding layer of the student model is initialized using the pre-trained embeddings from the MobileBERT teacher model:

\begin{equation}
E_{\text{student}} = E_{\text{teacher}} \in \mathbb{R}^{V \times d}
\label{eq:embedding_equality}
\end{equation}

Here, $V$ denotes the vocabulary size (approximately 30,000 tokens), and $d = 128$ is the embedding dimension. The student model builds upon the earlier multi-head attention BiLSTM architecture, with the following configuration adapted for knowledge distillation: 4 attention heads (key dimension 64), BiLSTM hidden size 128, and a global average pooling layer followed by a dropout and a dense projection. The multi-head attention computations follow equations \eqref{eq:multihead_attention} to \eqref{eq:final_dense_classification}.

\subsubsection{Training Protocol and Optimization}
The distillation training process follows a structured protocol designed to maximize knowledge transfer while minimizing overfitting. As described in Algorithm \ref{alg:distillation}, the MobileBERT model is first fine-tuned on the phishing detection task and then kept frozen during student training. The student is trained using the Adam optimizer with a learning rate of $1e$-$4$, a batch size of 32, and a total of 3 epochs. Each training step integrates both hard supervision via cross-entropy loss with ground-truth labels ($L_{\text{hard}}$) and soft supervision via Kullback-Leibler divergence ($L_{\text{soft}}$) between the softened teacher and student outputs. The combined loss incorporates a temperature scaling factor to ensure effective gradient flow, enabling the student to replicate the teacher's semantic behavior efficiently.

\begin{algorithm}
\caption{Knowledge Distillation Process}
\label{alg:distillation}
\KwIn{
    Training dataset $D$;
    Teacher model $f_{\text{teacher}}$ (MobileBERT, fine-tuned and frozen); Student model $f_{\text{student}}$ (BiLSTM + Multi-Head Attention); Hard loss weight $\alpha = 0.5$; Temperature $\tau = 2.0$; Learning rate $\eta = 1e\text{-}4$; Epochs $E = 3$; Batch size $B = 32$; }

\KwOut{Trained student model $f_{\text{student}}$}

Freeze parameters of $f_{\text{teacher}}$\;

Initialize Adam optimizer with learning rate $\eta$ for $f_{\text{student}}$\;

\For{$epoch \gets 1$ \KwTo $E$}{
    \ForEach{batch $(X_{batch}, Y_{batch})$ in $D$}{
        
        \tcp{Teacher forward pass}
        $Z_T \gets f_{\text{teacher}}(X_{batch})$\;
        $P_T^{\text{soft}} \gets \text{softmax}(Z_T / \tau)$\;
        
        \tcp{Student forward pass}
        $Z_S \gets f_{\text{student}}(X_{batch})$\;
        $P_S^{\text{soft}} \gets \text{softmax}(Z_S / \tau)$\;
        
        \tcp{Hard loss}
        
        $L_{\text{hard}} \gets \text{CrossEntropy}(Y_{batch}, \text{softmax}(Z_S))$\; 
        
        \tcp{Soft loss}
        $L_{\text{soft}} \gets \text{KL\_Divergence}(P_T^{\text{soft}} \,\|\, P_S^{\text{soft}})$\;
        
        \tcp{Combined loss}
        $L_{\text{distill}} \gets \alpha \cdot L_{\text{hard}} + (1 - \alpha) \cdot \tau^2 \cdot L_{\text{soft}}$\;
        
        \tcp{Backpropagation and update}
        $L_{\text{distill}}.\text{backward}()$\;
        
        optimizer.step()\;
        
        optimizer.zero\_grad()\;
    }
}
\Return{$f_{\text{student}}$}
\end{algorithm}

%% file: evaluation.tex
\label{evaluation}

\begin{figure*}[!t]
    \centering
    \includegraphics[width=\textwidth]{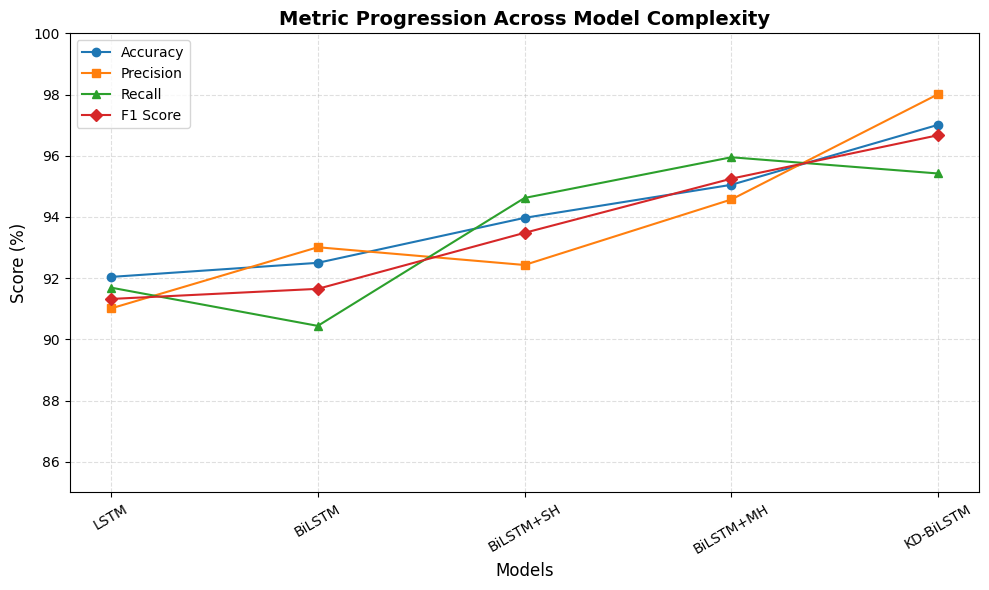}
    \caption{Metric progression across model complexity. Accuracy, Precision, Recall, and F1 score improve consistently from LSTM to BiLSTM variants, peaking with the KD-BiLSTM model.}
    \label{fig:model_progression_overview}
\end{figure*}

In this section, we present a comprehensive evaluation of the proposed phishing detection models. We systematically assess the performance of various configurations, including baseline BiLSTM (with and without attention), Multi-Head Attention BiLSTM, and knowledge-distilled BiLSTM models. To further evaluate robustness and generalization, we conduct cross-distribution experiments where models are trained and tested across different combinations of real-world (original) and synthetically generated (LLM-based) phishing emails. As shown in Figure~\ref{fig:model_progression_overview}, model performance improves steadily with architectural enhancements and knowledge distillation.

The figure ~\ref{fig:model_progression_overview} shows the progressive performance gains across model enhancements. The integration of multi-head attention and knowledge distillation leads to significant improvements in all metrics, which shows the impact of architectural and semantic enhancements on phishing email detection.

All experiments were conducted on Google Colab using L4 compute instances equipped with \SI{22.5}{\giga\byte} of GPU memory. To ensure consistency and minimize variance due to random initialization, we employed a 5-fold stratified cross-validation procedure across all classification models. In each fold, the dataset was partitioned into 72\% for training, 8\% for validation, and 20\% for testing.

\subsection{Evaluation Metrics}

The evaluation of model performance used standard binary classification metrics which are appropriate for phishing detection because class imbalance occurs frequently in this domain. The confusion matrix from each experiment tracked classification results through True Positives (TP), True Negatives (TN), False Positives (FP) and False Negatives (FN) as presented in Table~\ref{binaryconfusion}.

\begin{table}[H]
\caption{Confusion matrix for phishing email detection.}
\centering
\label{binaryconfusion}
\begin{tabular}{llll}
\multicolumn{2}{c}{\multirow{2}{*}{}} & \multicolumn{2}{c}{\tiny Actual} \\ \cline{3-4}
\multicolumn{2}{c}{} & \tiny Phishing (Positive) & \tiny Legitimate (Negative) \\ \hline
\multirow{2}{*}{\tiny Predicted} & \tiny Phishing (Positive) & \tiny TP & \tiny FP \\
 & \tiny Legitimate (Negative) & \tiny FN & \tiny TN \\ \hline
\end{tabular}
\end{table}

From the confusion matrix, the following metrics are calculated:

\begin{equation}
\text{Accuracy} = \frac{TP + TN}{TP + TN + FP + FN} 
\end{equation}
\begin{equation}
\text{Precision} = \frac{TP}{TP + FP}
\end{equation}
\begin{equation}
\text{Recall} = \frac{TP}{TP + FN} 
\end{equation}

\begin{equation}
\text{F1\ Score} = \frac{2 \times \text{Precision} \times \text{Recall}}{\text{Precision} + \text{Recall}}
\end{equation}

The overall correctness score of Accuracy can be misleading when data is imbalanced. Precision measures the proportion of correctly identified malicious phishing emails among all predicted phishing emails while Recall measures the proportion of detected phishing attempts. The F1 Score provides a balanced evaluation of both Precision and Recall which makes it crucial for assessing models when class distributions are uneven.

\begin{table*}[!t]
    \setlength{\tabcolsep}{1pt}
    \centering
    \caption{Average Performance results for Baseline Models LSTM and Bi-LSTM Using 5-Fold Cross-Validation}
    \label{tab:LSTM_BiLSTM_table}
    \renewcommand{\arraystretch}{1.3} 
\begin{tabular}{>{\centering}p{0.13\linewidth}>{\centering}p{0.13\linewidth}>{\centering}p{0.1\linewidth}>{\centering}p{0.08\linewidth}>{\centering}p{0.07\linewidth}>{\centering}p{0.09\linewidth}>{\centering}p{0.1\linewidth}>{\centering}p{0.08\linewidth}>{\centering}p{0.08\linewidth}p{0.08\linewidth}}  
     \bottomrule
 \rowcolor[HTML]{C0C0C0}  \bfseries \scriptsize Classifier Model 
  \space & \bfseries \space \scriptsize Training and Testing Scenario
\space & \bfseries \space \scriptsize No of Parameters
\centering \space & \bfseries \space \scriptsize Train Time (s) 
 \space & \bfseries \space \scriptsize Test Time (s) 
 \space & \bfseries \space \scriptsize Total Samples

\space& \centering \bfseries \scriptsize \space ACC  \tiny (Avg.) \space & \centering \bfseries \space \scriptsize P  \tiny (Avg.) \space & \centering \bfseries \space \scriptsize R \tiny (Avg.)   \space & \bfseries \scriptsize \space $\boldsymbol{F_1}$  \tiny (Avg.) \space \\  \hline
  
\rowcolor[HTML]{F0F0F0} \scriptsize LSTM  & \scriptsize Orig-Orig & \scriptsize 16,961 &  \scriptsize 9.25 &  \scriptsize 0.56 & \scriptsize 13,692 & \scriptsize 86.96  & \scriptsize 86.88 & \scriptsize 84.42 & \scriptsize 85.41 \\  \cline{2-10}

\bfseries \scriptsize   & \scriptsize Gen-Gen & \scriptsize 16,961  &  \scriptsize 9.30 &  \scriptsize 0.56 & \scriptsize 13,692 & \scriptsize 88.94 & \scriptsize 86.45 & \scriptsize 90.26 & \scriptsize 88.21 \\  \cline{2-10}
\bfseries \scriptsize   & \scriptsize Orig-Gen & \scriptsize 16,961 &  \scriptsize 9.26 &  \scriptsize 0.55 & \scriptsize 13,692 & \scriptsize 80.70  & \scriptsize \textcolor{red}{71.82} & \scriptsize 96.46 & \scriptsize 82.17  \\  \cline{2-10}

\bfseries \scriptsize   & \scriptsize Gen-Orig & \scriptsize 16,961 &  \scriptsize 9.23 &  \scriptsize 0.56 & \scriptsize 13,692 & \scriptsize 71.89 & \scriptsize 88.14 & \scriptsize \textcolor{red}{45.10} & \scriptsize  \textcolor{red}{58.97}  \\  \cline{2-10}

\rowcolor[HTML]{E6F7FF} \bfseries \scriptsize   & \scriptsize Mixture of all & \scriptsize 16,961 &  \scriptsize 13.98 &  \scriptsize 0.83  & \scriptsize 27,384 & \scriptsize 92.04  & \scriptsize 91.01  & \scriptsize 91.69 & \scriptsize 91.32 \\  \hline 

\rowcolor[HTML]{F0F0F0} \scriptsize Bi-LSTM  & \scriptsize Orig-Orig & \scriptsize 33,921 &  \scriptsize 13.07  &  \scriptsize 0.88 & \scriptsize 13,692 & \scriptsize 90.47  & \scriptsize 89.76 & \scriptsize 89.29  & \scriptsize 89.52  \\  \cline{2-10}

\bfseries \scriptsize   & \scriptsize Gen-Gen & \scriptsize 33,921 &  \scriptsize 13.14 &  \scriptsize 0.87 & \scriptsize 13,692  & \scriptsize 90.72  & \scriptsize 91.60 & \scriptsize 87.73 & \scriptsize 89.57  \\  \cline{2-10}

\bfseries \scriptsize   & \scriptsize Orig-Gen & \scriptsize 33,921 &  \scriptsize 13.08 &  \scriptsize 0.86 & \scriptsize 13,692 & \scriptsize 77.57  & \scriptsize 67.90 & \scriptsize 97.44 & \scriptsize 79.93  \\  \cline{2-10}

\bfseries \scriptsize   & \scriptsize Gen-Orig & \scriptsize 33,921 &  \scriptsize 13.00  &  \scriptsize 0.87 & \scriptsize 13,692 & \scriptsize 75.64 & \scriptsize 91.97 & \scriptsize  \textcolor{red}{51.19} & \scriptsize  \textcolor{red}{65.59} \\  \cline{2-10}

\rowcolor[HTML]{E6F7FF} \bfseries \scriptsize   & \scriptsize Mixture of all & \scriptsize 33,921 &  \scriptsize 21.43 &  \scriptsize 1.30 & \scriptsize 27,384 & \scriptsize 92.50  & \scriptsize 93.01  & \scriptsize 90.44 & \scriptsize 91.65  \\  
    \bottomrule
    \end{tabular}
    \label{Table_Results_LSTM-BiLSTM}
    \begin{tablenotes}
    \item \scriptsize \textbf{Note:} ACC = Accuracy, P = Precision, R = Recall, F1 = F1 Score.
    The BiLSTM model performs better than LSTM in every evaluation scenario with the biggest performance differences occurring during cross-distribution assessments. The results confirm that bidirectional modeling provides better performance for detecting phishing emails because it can handle wider contextual information.
\end{tablenotes}
\end{table*}

\subsection{Evaluation Results for the Baseline LSTM and BiLSTM Models}
This subsection presents the performance results of the baseline LSTM and BiLSTM models to demonstrate the stepwise improvements of the proposed architecture. As summarized in Tables \ref{Table_Results_LSTM-BiLSTM}, \ref{Table_Results_Attention_Mechanism}, and \ref{Table_Results_Knowledge_Distillation}, each model configuration was evaluated across four training-testing scenarios:

\begin{itemize}
    \item \textbf{Orig-Orig:} Training and testing are performed exclusively on real-world (original) phishing and legitimate emails.
    
    \item \textbf{Gen-Gen:} Both training and testing are conducted using LLM-generated phishing samples and corresponding legitimate emails.
    
    \item \textbf{Orig-Gen:} The model is trained on real-world data and tested on LLM-generated phishing samples to assess its ability to generalize to synthetic attacks.
    
    \item \textbf{Gen-Orig:} The model is trained on LLM-generated data and tested on real-world samples to evaluate reverse generalizability.
\end{itemize}
These scenarios are designed to evaluate both in-distribution (Orig-Orig, Gen-Gen) and cross-distribution (Orig-Gen, Gen-Orig) performance. They reveal the advantages of using synthetic data during training and expose the challenges caused by distributional shifts at inference time. The results of the LSTM configuration, shown in the first part of Table \ref{Table_Results_LSTM-BiLSTM}, and the evaluation results of BiLSTM model are presented in the second part of the same table.


\begin{table*}[!t]
    \setlength{\tabcolsep}{1pt}
    \centering
    \caption{Average Performance of BiLSTM Models with Attention Mechanisms Using 5-Fold Cross-Validation}
    \label{tab:BiLSTMSH_BiLSTMMH_table}
    \renewcommand{\arraystretch}{1.3} 
\begin{tabular}{>{\centering}p{0.2\linewidth}>{\centering}p{0.13\linewidth}>{\centering}p{0.1\linewidth}>{\centering}p{0.08\linewidth}>{\centering}p{0.07\linewidth}>{\centering}p{0.09\linewidth}>{\centering}p{0.07\linewidth}>{\centering}p{0.07\linewidth}>{\centering}p{0.07\linewidth}p{0.07\linewidth}}  
     \bottomrule
 \rowcolor[HTML]{C0C0C0}  \bfseries \scriptsize Classifier Model 
  \space & \bfseries \space \scriptsize Training and Testing Scenario
\space & \bfseries \space \scriptsize No of Parameters
\centering \space & \bfseries \space \scriptsize Train Time (s) 
 \space & \bfseries \space \scriptsize Test Time (s) 
 \space & \bfseries \space \scriptsize Total Samples

\space& \centering \bfseries \scriptsize \space ACC  \tiny (Avg.) \space & \centering \bfseries \space \scriptsize P  \tiny (Avg.) \space & \centering \bfseries \space \scriptsize R \tiny (Avg.)   \space & \bfseries \scriptsize \space $\boldsymbol{F_1}$  \tiny (Avg.) \space \\  \hline
  
\rowcolor[HTML]{F0F0F0} \scriptsize Bi-LSTM with Single-head attention  & \scriptsize Orig-Orig & \scriptsize 66,561 &  \scriptsize 17.07 &  \scriptsize  0.95 & \scriptsize 13,692 & \scriptsize 92.52  & \scriptsize 91.25 &\scriptsize  92.57 & \scriptsize  91.87 \\  \cline{2-10}

\bfseries \scriptsize   & \scriptsize Gen-Gen & \scriptsize 66,561 &  \scriptsize 16.31 &  \scriptsize 0.94  & \scriptsize 13,692 & \scriptsize  89.02 & \scriptsize 86.80 &\scriptsize 89.75 & \scriptsize 88.23  \\  \cline{2-10}

\bfseries \scriptsize   & \scriptsize Orig-Gen & \scriptsize 66,561 &  \scriptsize 16.28 &  \scriptsize  0.94 & \scriptsize 13,692 & \scriptsize 84.65  & \scriptsize 77.20 &\scriptsize 95.02 & \scriptsize 85.03  \\  \cline{2-10}

\bfseries \scriptsize   & \scriptsize Gen-Orig & \scriptsize 66,561 &  \scriptsize 16.28 &  \scriptsize  0.94  & \scriptsize 13,692 & \scriptsize  \textcolor{red}{74.32} & \scriptsize 90.07 &\scriptsize \textcolor{red}{49.09} & \scriptsize \textcolor{red}{63.05}  \\  \cline{2-10}

\rowcolor[HTML]{E6F7FF} \bfseries \scriptsize   & \scriptsize Mixture of all & \scriptsize 66,561 &  \scriptsize 26.26 &  \scriptsize 1.37 & \scriptsize 27,384 & \scriptsize  93.97 & \scriptsize 92.43 &\scriptsize 94.62 & \scriptsize  93.48 \\  \hline 

\rowcolor[HTML]{F0F0F0} \scriptsize Bi-LSTM with Multi-head attention  & \scriptsize Orig-Orig & \scriptsize 494,977 &  \scriptsize 45.74 &  \scriptsize 1.07  & \scriptsize 13,692 & \scriptsize  94.46 & \scriptsize 93.10 &\scriptsize 94.96  & \scriptsize 94.57  \\  \cline{2-10}

\bfseries \scriptsize   & \scriptsize Gen-Gen & \scriptsize 494,977 &  \scriptsize 45.97 &  \scriptsize 1.07  & \scriptsize 13,692 & \scriptsize  94.19 & \scriptsize 92.44 &\scriptsize 95.04 & \scriptsize 93.72  \\  \cline{2-10}

\bfseries \scriptsize   & \scriptsize Orig-Gen & \scriptsize 494,977 &  \scriptsize 46.53 &  \scriptsize 1.08 & \scriptsize 13,692 & \scriptsize 91.84  & \scriptsize 89.20 &\scriptsize 95.84 & \scriptsize 92.40  \\  \cline{2-10}

\bfseries \scriptsize   & \scriptsize Gen-Orig & \scriptsize 494,977 &  \scriptsize 45.77 &  \scriptsize 1.07 & \scriptsize 13,692 & \scriptsize \textcolor{blue}{79.78}  & \scriptsize 89.64 &\scriptsize \textcolor{blue}{62.97} & \scriptsize \textcolor{blue}{73.82}  \\  \cline{2-10}

\rowcolor[HTML]{E6F7FF} \bfseries \scriptsize   & \scriptsize Mixture of all & \scriptsize 494,977 &  \scriptsize 84.02 &  \scriptsize 1.51 & \scriptsize 27,384 & \scriptsize 95.05  & \scriptsize 94.57 &\scriptsize 95.95 & \scriptsize  95.25 \\  
 
    \bottomrule
    \end{tabular}
    \label{Table_Results_Attention_Mechanism}
    \begin{tablenotes}
    \item \scriptsize \textbf{Note:} ACC = Accuracy, P = Precision, R = Recall, F1 = F1 Score.
   The table presents performance metrics between BiLSTM models that use single-head and multi-head attention mechanisms. The multi-head attention model demonstrates steady performance gains especially during cross-distribution testing because its parallel attention heads enhance both generalization and feature extraction capabilities.
\end{tablenotes}

\end{table*}%


\subsection{Evaluation Results for BiLSTM Models with Single-head and Multi-head Attention }

Extending the BiLSTM architecture, this set of experiments incorporates single-head and multi-head attention mechanisms to emphasize tokens with higher semantic relevance to phishing indicators. By prioritizing critical phrases, the attention layer enhances the model’s discrimination between phishing and legitimate messages. As shown in the first part of Table \ref{Table_Results_Attention_Mechanism}, the single-head attention configuration improves precision and recall, particularly in scenarios involving LLM-generated samples. The second part of Table \ref{Table_Results_Attention_Mechanism} shows that the multi-head attention model further enhances performance, especially in cross-distribution settings. The observed gains in precision and F1 score demonstrate improved generalization and robustness, as the model benefits from attending to multiple phishing-relevant patterns in parallel.

\subsection{BiLSTM (with Multi-head attention and knowledge distillation from MobileBERT)}

\begin{table*}[!t]
    \setlength{\tabcolsep}{1pt}
    \centering
    \caption{Average Performance for the Student Model empowered by Knowledge Distillation Using 5-Fold Cross-Validation }
    \label{tab:KD-BiLSTMvsMobileBERT_table}
    \renewcommand{\arraystretch}{1.5} 
\begin{tabular}{>{\centering}p{0.25\linewidth}>{\centering}p{0.13\linewidth}>{\centering}p{0.1\linewidth}>{\centering}p{0.08\linewidth}>{\centering}p{0.07\linewidth}>{\centering}p{0.07\linewidth}>{\centering}p{0.07\linewidth}>{\centering}p{0.06\linewidth}>{\centering}p{0.06\linewidth}p{0.06\linewidth}}  
     \bottomrule
 \rowcolor[HTML]{C0C0C0}  \bfseries \scriptsize Classifier Model 
  \space & \bfseries \space \scriptsize Training and Testing Scenario
\space & \bfseries \space \scriptsize No of Parameters
\centering \space & \bfseries \space \scriptsize Train Time (s) 
 \space & \bfseries \space \scriptsize Test Time (s) 
 \space & \bfseries \space \scriptsize Total Samples

\space& \centering \bfseries \scriptsize \space ACC  \tiny (Avg.) \space & \centering \bfseries \space \scriptsize P  \tiny (Avg.) \space & \centering \bfseries \space \scriptsize R \tiny (Avg.)   \space & \bfseries \scriptsize \space $\boldsymbol{F_1}$  \tiny (Avg.) \space \\  \hline  
\rowcolor[HTML]{F0F0F0}  \scriptsize Bi-LSTM (Multi-head attention) with knowledge distillation  & \scriptsize Orig-Orig & \scriptsize 4.5 M &  \scriptsize 665.01 &  \scriptsize 3.3  & \scriptsize 13,692 & \scriptsize 97.58  & \scriptsize 97.48 &\scriptsize  97.21 & \scriptsize  97.34 \\  \cline{2-10}

\bfseries \scriptsize   & \scriptsize Gen-Gen & \scriptsize 4.5 M &  \scriptsize 667 &  \scriptsize 3.04 & \scriptsize 13,692  & \scriptsize 95.76 & \scriptsize 95.04 &\scriptsize  95.84 & \scriptsize  95.39  \\  \cline{2-10}

\bfseries \scriptsize   & \scriptsize Orig-Gen & \scriptsize 4.5 M &  \scriptsize 665.32 &  \scriptsize 2.97 & \scriptsize 13,692  & \scriptsize 89.57  & \scriptsize 82.33 &\scriptsize \textcolor{blue}{98.33} & \scriptsize  89.60  \\  \cline{2-10}

\bfseries \scriptsize   & \scriptsize Gen-Orig & \scriptsize 4.5 M &  \scriptsize 663.70 &  \scriptsize 2.95 & \scriptsize 13,692  & \scriptsize 90.05  & \scriptsize \textcolor{red}{83.12} &\scriptsize \textcolor{blue}{98.24} & \scriptsize 90.02  \\  \cline{2-10}

\rowcolor[HTML]{E6F7FF} \bfseries \scriptsize   & \scriptsize Mixture of all & \scriptsize \textcolor{blue}{4.5 M} &  \scriptsize \textcolor{blue}{1301.72} &  \scriptsize \textcolor{blue}{6.06} & \bfseries \scriptsize 27,384 & \scriptsize \textcolor{blue}{97.01}  & \scriptsize \textcolor{blue}{98.01} &\scriptsize \textcolor{blue}{95.42} & \scriptsize \textcolor{blue}{96.67} \\  \hline 

\rowcolor[HTML]{F0F0F0}  \scriptsize MobileBERT (for comparison)  & \scriptsize Orig-Orig & \scriptsize 25.3 M &  \scriptsize 1,651 &   \scriptsize 21  & \scriptsize 13,692 & \scriptsize 98.85 & \scriptsize 98.85 &\scriptsize  98.85 & \scriptsize  98.85  \\  \cline{2-10}

\bfseries \scriptsize   & \scriptsize Gen-Gen & \scriptsize 25.3 M  &  \scriptsize 1,516 &  \scriptsize 21 & \scriptsize 13,692 & \scriptsize 97.33 & \scriptsize 97.33 &\scriptsize  97.33 & \scriptsize  97.33  \\  \cline{2-10}

\bfseries \scriptsize   & \scriptsize Orig-Gen & \scriptsize 25.3 M  &  \scriptsize 1,527 &  \scriptsize 21 & \scriptsize 13,692 & \scriptsize 91.79 & \scriptsize 92.78 &\scriptsize  \textcolor{red}{91.79} & \scriptsize  91.80  \\  \cline{2-10}

\bfseries \scriptsize   & \scriptsize Gen-Orig & \scriptsize 25.3 M  &  \scriptsize 1,517 &  \scriptsize 21 & \scriptsize 13,692 & \scriptsize 91.79 & \scriptsize \textcolor{blue}{92.78} &\scriptsize \textcolor{red}{91.79} & \scriptsize 91.80  \\  \cline{2-10}

\rowcolor[HTML]{E6F7FF} \bfseries \scriptsize   & \scriptsize Mixture of all & \scriptsize \textcolor{red}{25.3 M}  &  \scriptsize \textcolor{red}{3,020} &  \scriptsize \textcolor{red}{42} & \bfseries \scriptsize 27,384 & \scriptsize \textcolor{blue}{98.56} & \scriptsize \textcolor{blue}{98.56} &\scriptsize \textcolor{blue}{98.56} & \scriptsize \textcolor{blue}{98.56} \\  
 
    \bottomrule
    \end{tabular}
    \label{Table_Results_Knowledge_Distillation}
    \begin{tablenotes}
    \item \scriptsize \textbf{Note:} ACC = Accuracy, P = Precision, R = Recall, F1 = F1 Score, M = Million.
   The KD-BiLSTM model demonstrates comparable performance to MobileBERT through its reduced parameter count and accelerated training and inference speeds. The model's performance characteristics make it suitable for deployment on edge devices.

\end{tablenotes}

\end{table*}

\begin{table*}[!t]
\centering
\scriptsize
\setlength{\tabcolsep}{3pt}
\renewcommand{\arraystretch}{1.12}
\begin{threeparttable}
\caption{ACC and weighted F1 with runtime and params for four single-source splits and the Mixture-of-all. The proposed KD--BiLSTM + Multi-Head Attention (4.5M params) is within $\sim$1--2.3 F1 points of transformer baselines on Mixture-of-all while cutting inference time by $\sim$5--19\(\times\) and model size by $\sim$20--800\(\times\). Means are over three seeds where available; single-run cases are noted in the text.}
\label{tab:avg_scenarios}
\begin{tabularx}{\textwidth}{@{} >{\raggedright\arraybackslash}X >{\raggedright\arraybackslash}p{2.35cm} c r r r r @{}}
\hline
\rowcolor[HTML]{C0C0C0} \textbf{Classifier Model} & \textbf{Training / Testing Scenario} & \textbf{Params (M)} & \textbf{Train (s)} & \textbf{Test (s)} & \textbf{ACC} & \textbf{F1} \\
\hline

\rowcolor{black!6}\multicolumn{7}{@{}l}{\textbf{DeBERTaV3 (base)}}\\
 & Orig--Orig & 86  & 1996.85 & 22.71 & 99.25 & 99.25 \\
 & Gen--Gen   & 86  & 1847.81 & 22.77 & 97.75 & 97.76 \\
 & Orig--Gen  & 86  & 1851.04 & 22.70 & 88.85 & 88.65 \\
 & Gen--Orig  & 86  & 1851.82 & 22.72 & 93.27 & 93.26 \\
\rowcolor{blue!6}
 & Mixture of all & 86 & 3532.36 & 38.34 & 98.76 & 98.75 \\
\addlinespace[3pt]

\rowcolor{black!6}\multicolumn{7}{@{}l}{\textbf{DeepSeek-R1 Distill Qwen-1.5B (cls head)}}\\
 & Orig--Orig & 1780  & 4615.88 & 29.27 & \textbf{99.34} & \textbf{99.34} \\
 & Gen--Gen   & 1780  & 4611.16 & 29.31 & 97.54 & 97.50 \\
 & Orig--Gen  & 1780  & 4610.98 & 29.30 & 92.81 & 92.78 \\
 & Gen--Orig  & 1780  & 4611.80 & 29.30 & \textbf{94.22} & \textbf{94.22} \\
\rowcolor{blue!6}
 & Mixture of all & 1780 & 8872.66 & 56.40 & 98.84 & 98.83 \\
\addlinespace[3pt]

\rowcolor{black!6}\multicolumn{7}{@{}l}{\textbf{Microsoft Phi-4 mini instruct (cls head)} }\\
 & Orig--Orig & 3840  & 16093.65 & 60.13 & 99.13 & 99.11 \\
 & Gen--Gen   & 3840  & 16383.63 & 60.55 & \textbf{97.98} & \textbf{97.97} \\
 & Orig--Gen  & 3840  & 16203.54 & 60.70 & 91.24 & 91.24 \\
 & Gen--Orig  & 3840  & 16395.91 & 60.51 & 90.23 & 90.23 \\
\rowcolor{blue!6}
 & Mixture of all & 3840 & 31246.12 & 116.65 & 96.48 & 96.45 \\
\addlinespace[3pt]

\rowcolor{black!6}\multicolumn{7}{@{}l}{\textbf{T5-base (cls head)}}\\
 & Orig--Orig & 109.6 & 1363.53 & 16.34 & 98.91 & 98.91 \\
 & Gen--Gen   & 109.6 & 1364.45 & 16.34 & 97.83 & 97.83 \\
 & Orig--Gen  & 109.6 & 1363.97 & 16.34 & \textbf{92.99} & \textbf{92.99} \\
 & Gen--Orig  & 109.6 & 1364.46 & 16.35 & 93.47 & 93.47 \\
\rowcolor{blue!6}
 & Mixture of all & 109.6 & 2624.86 & 31.43 & 98.73 & 98.73 \\
\addlinespace[3pt]

\rowcolor{black!6}\multicolumn{7}{@{}l}{\textbf{ModernBERT-base}}\\
 & Orig--Orig & 149 & 1604.08 & 19.39 & 99.17 & 99.17 \\
 & Gen--Gen   & 149  & 1590.79 & 19.40 & 97.75 & 97.76 \\
 & Orig--Gen  & 149  & 1591.87 & 19.41 & 87.31 & 87.04 \\
 & Gen--Orig  & 149  & 1596.41 & 19.39 & 94.04 & 94.03 \\
\rowcolor{blue!6}
 & Mixture of all & 149 & 3037.06 & 37.26 & \textbf{98.98} & \textbf{98.98} \\
\addlinespace[3pt]

\rowcolor{black!6}\multicolumn{7}{@{}l}{\textbf{Proposed: KD-BiLSTM + Multi-Head Attention (student distilled from MobileBERT)}}\\
 & Orig--Orig & \textbf{4.5} & \textbf{665.01} & \textbf{3.3} & 97.58 & 97.34 \\
 & Gen--Gen   & \textbf{4.5} & \textbf{667} & \textbf{3.04} & 95.76 & 95.39 \\
 & Orig--Gen  & \textbf{4.5} & \textbf{665.32} & \textbf{2.97} & 89.57 & 89.60 \\
 & Gen--Orig  & \textbf{4.5} & \textbf{663.70} & \textbf{2.95} & 90.05 & 90.02 \\
\rowcolor{blue!6}
 & Mixture of all & \textbf{4.5} & \textbf{1301.72} & \textbf{6.06} & 97.01 & 96.67\\
\bottomrule
\end{tabularx}

\begin{tablenotes}[flushleft]
\footnotesize
\item \textbf{Notes:} ACC $=$ test accuracy (\%). F1 $=$ weighted F1 (\%). Totals are train+val+test; single-scenario $=$ 13{,}692; Mixture $=$ 27{,}384.
\end{tablenotes}
\end{threeparttable}
\end{table*}



\paragraph{Reading Table~\ref{tab:avg_scenarios}}
Table~\ref{tab:avg_scenarios}  provides an overview of accuracy and efficiency for all models and splits. Three patterns stand out.
\textbf{(i) Cross-distribution stress:} All models show a dip on \textit{Orig--Gen}/\textit{Gen--Orig}, indicating that LLM-rewrites are harder than in-distribution cases, and motivating training with mixed sources. 
\textbf{(ii) KD-BiLSTM’s edge profile:} Despite having only 4.5M parameters, KD-BiLSTM stays within $\approx$1–2.3 F1 points of the stronger transformer baselines on \textit{Mixture-of-all}, while offering markedly lower training and inference time. 
\textbf{(iii) Deployment relevance:} For latency-sensitive email pipelines, Model size and \textit{Test (s)} matter as much as raw F1. KD-BiLSTM’s small footprint and fast inference therefore provide a practical path to inline or on-device filtering with minimal accuracy trade-off.

The evaluation of Mixture of all demonstrates its practical advantages. The KD-BiLSTM + Multi-Head Attention model achieves a weighted F1 score of 96.67 with 6.06 seconds of test time and 4.5 million parameters while ModernBERT-base reaches 98.98/37.26 s/149 M and T5-base reaches 98.73/31.43 s/109.6 M and DeBERTaV3-base reaches 98.75/38.34 s/86 M and DeepSeek-R1 (cls head) reaches 98.83/56.40 s/1780 M and Phi-4 mini (cls head) reaches 96.45/116.65 s/3840 M. The results show that KD-BiLSTM stands as the only solution which achieves both fast inference and low memory usage while maintaining F1 scores that are close to state-of-the-art transformer models by 1–2.3 points which typically exceeds production email pipeline constraints related to latency and memory usage.

\subsection{Pareto view on the \textit{Mixture of all}}
\label{subsec:pareto-mixture}
The explicit representation of accuracy–latency–size trade-off requires us to evaluate our \textbf{KD-BiLSTM + Multi-Head Attention} against each baseline on the \emph{Mixture of all} split using radar charts (Fig.~\ref{fig:radar_grid_3x2}). The four axes are \emph{Train time}, \emph{Test time}, \emph{Params (M)}, and \emph{F1}. Time and size are inverted and normalized so that larger values are better on every axis.

\subsection{Discussion and Comparison}
The aggregate data from Table~\ref{tab:avg_scenarios} receives its context from the Pareto view presented in Figure~\ref{fig:radar_grid_3x2}. The F1 performance of KD-BiLSTM matches transformer leaders while maintaining lower latency and size requirements which appeal to in-path mail gateways and endpoint clients.
\subsubsection{\textbf{LSTM vs. BiLSTM}}
The experimental results from the baseline models (LSTM and BiLSTM) underscore the significance of both data diversity and architectural enhancements in improving phishing detection performance.  As shown in Table \ref{Table_Results_LSTM-BiLSTM}, models trained exclusively on LLM-generated phishing emails (Gen-Gen, Gen-Orig) achieve high precision (86.45\% and 88.14\% in the LSTM model), indicating their ability to minimize false positives. However, these models exhibit reduced recall (45.10\% in the LSTM Gen-Orig scenario), suggesting poor sensitivity to real phishing instances. This imbalance leads to a lower F1 score despite high precision, emphasizing that synthetic data alone cannot fully capture the variability of real-world attacks.



\begin{figure*}[!t]
  \centering
  \setlength{\tabcolsep}{3pt}

  \begin{tabular}{@{}c c@{}}
    \includegraphics[width=0.49\textwidth]{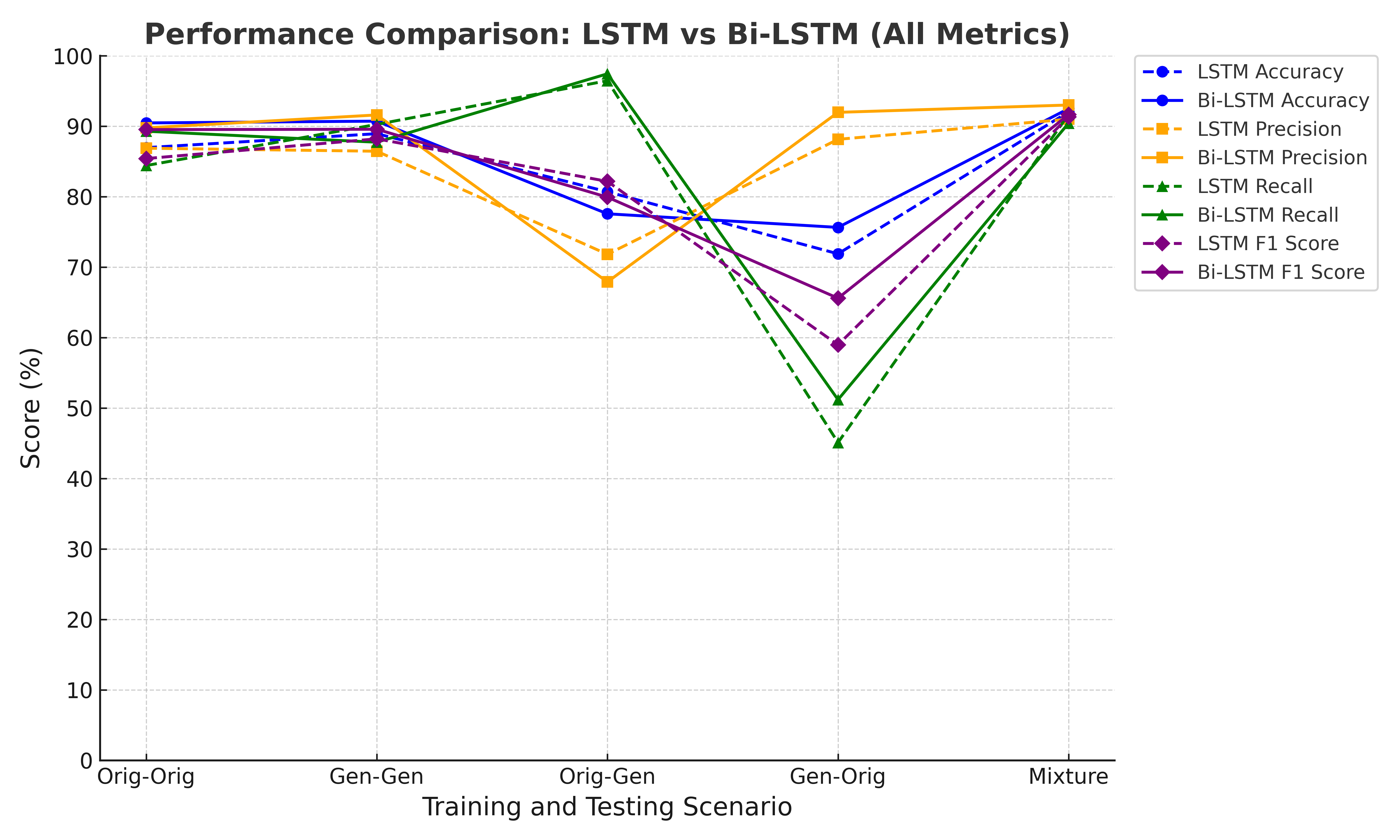} &
    \includegraphics[width=0.49\textwidth]{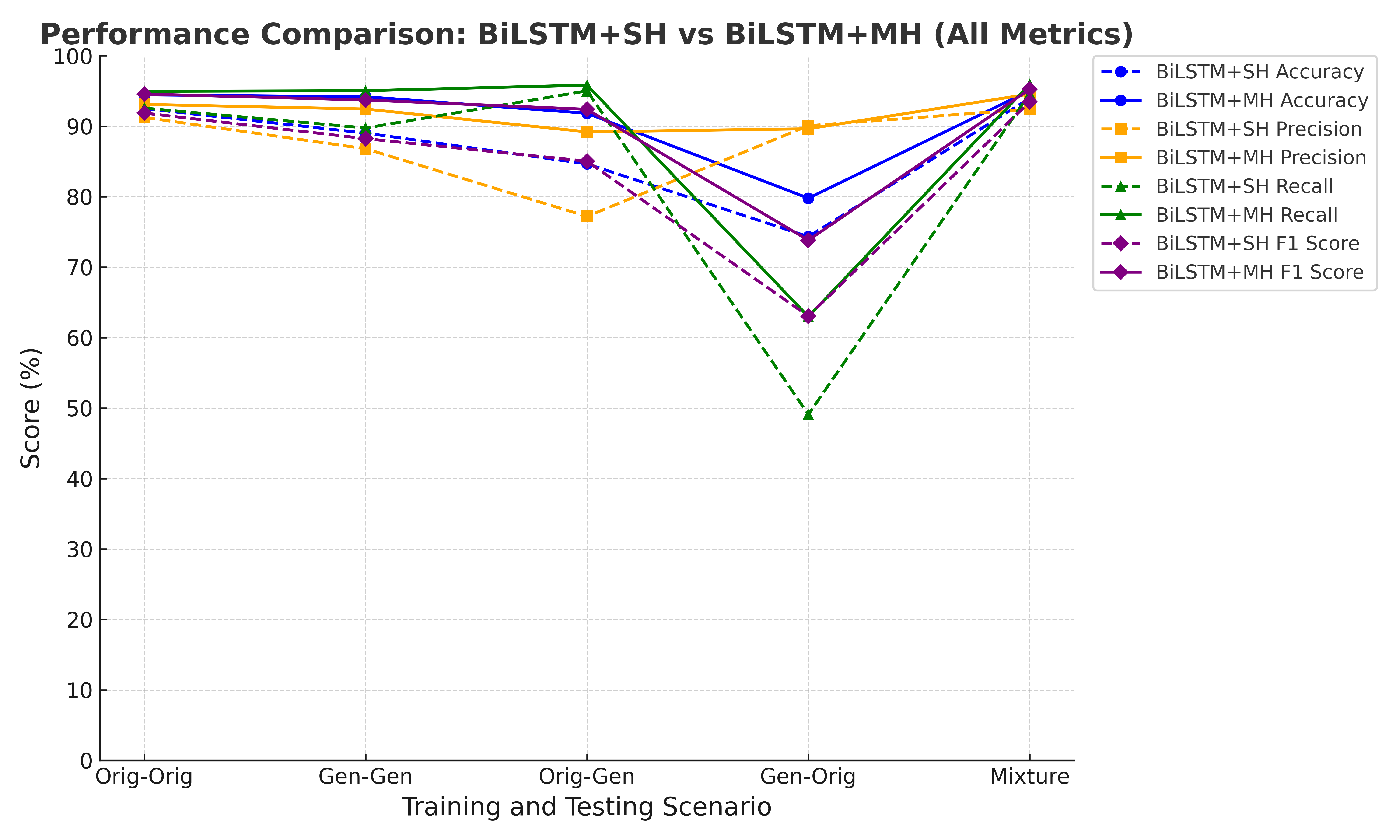} \\
    \footnotesize (a) LSTM vs.\ BiLSTM \label{fig:lstm_bilstm} &
    \footnotesize (b) Single-head vs.\ Multi-head \label{fig:sh_mh} \\
    \multicolumn{2}{c}{\vspace{4pt}} \\

    \multicolumn{2}{c}{\includegraphics[width=0.49\textwidth]{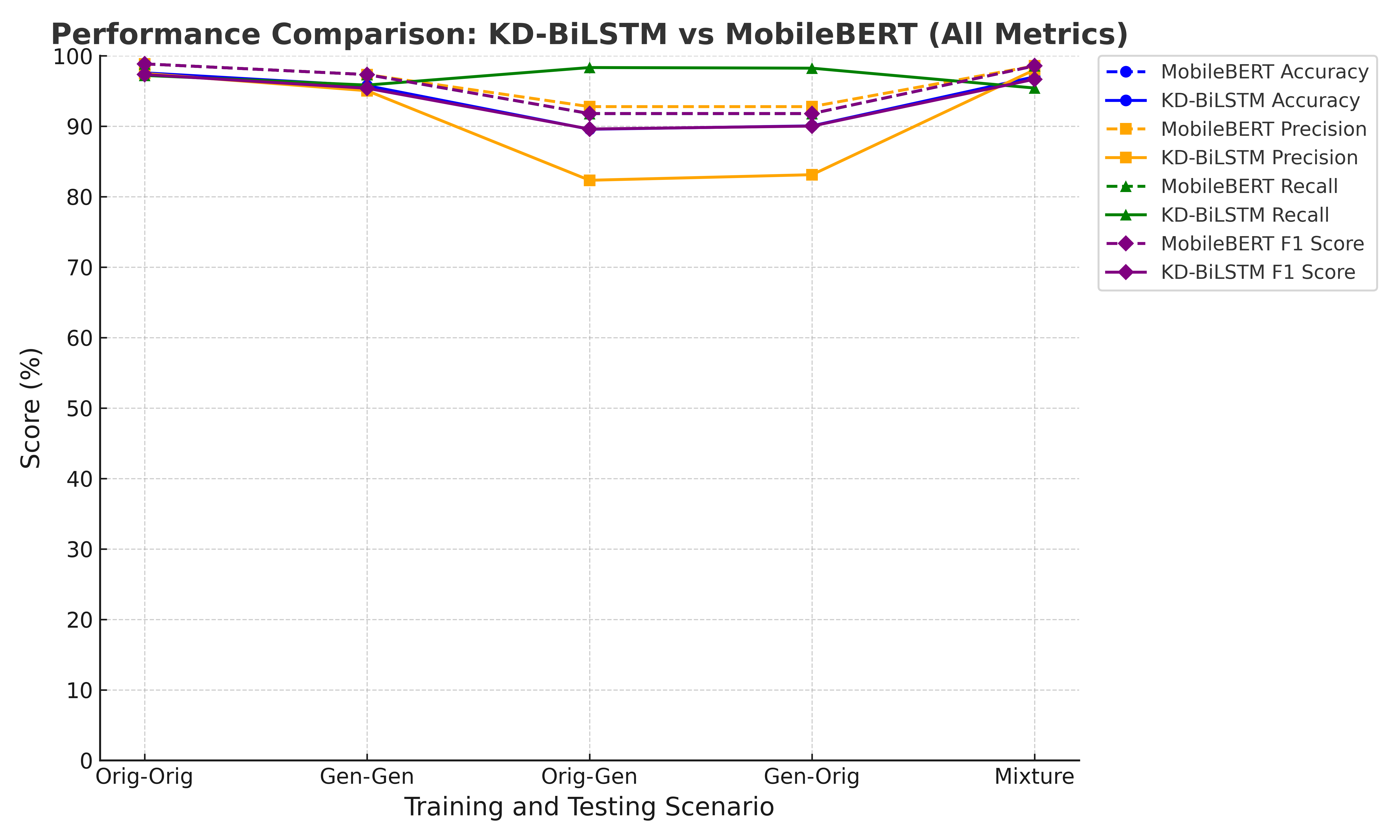}} \\
    \multicolumn{2}{c}{\footnotesize (c) KD-BiLSTM vs.\ MobileBERT \label{fig:kd_mobilebert}} \\
  \end{tabular}

  \caption{\textbf{Performance comparisons across evaluation scenarios.}
  (a) The Mixture configuration maintains high scores across metrics, while Gen--Orig yields the largest Recall drop when training on synthetic data and testing on real data.
  (b) Multi-head attention improves generalizability and maintains stable performance across evaluation scenarios.
  (c) KD-BiLSTM performs close to MobileBERT across scenarios at much lower cost, consistent with Table~\ref{tab:KD-BiLSTMvsMobileBERT_table}.}
  \label{fig:eval_threeplots_composite}
\end{figure*}





The results are clearly visible in Figure~\ref{fig:eval_threeplots_composite}(a), where the Gen-Orig scenario shows a significant drop (especially in Recall) across both LSTM and BiLSTM models. In contrast, the Mixture setup yields consistently strong results across all metrics, and BiLSTM (solid lines) remains above LSTM (dashed lines) throughout, confirming its superior ability to generalize in diverse data conditions.



In contrast, training on mixture of both real and synthetic data yields more balanced and robust performance. For instance, the BiLSTM model trained on the combined dataset achieves an accuracy of 92.50\%, precision of 93.01\%, recall of 90.44\%, and an F1 score of 91.65\%, outperforming all single-source training configurations. Additionally, the BiLSTM consistently outperforms the unidirectional LSTM across all evaluation scenarios, underscoring the advantage of incorporating bidirectional context for improved phishing detection.

\subsubsection{\textbf{Single-Head Attention vs. Multi-Head Attention}}

Table \ref{Table_Results_Attention_Mechanism} shows the benefit of integrating attention mechanisms into BiLSTM models for phishing email detection. Both single-head and multi-head attention improve the model’s ability to focus on phishing-relevant cues, but the multi-head attention configuration yields notably better generalization and performance across all evaluation scenarios.


The visual representation in Figure ~\ref{fig:eval_threeplots_composite}(b) supports the findings presented in Table ~\ref{tab:BiLSTMSH_BiLSTMMH_table} by showing the metric variations. The parallel attention mechanism of multi-head attention enables better recall performance particularly when data distribution shifts occur because it can process multiple semantic cues simultaneously.

While models trained exclusively on synthetic data still suffer from lower recall, the multi-head configuration demonstrates a meaningful improvement in sensitivity, achieving a recall of 62.97\% in the Gen-Orig setting, compared to 51.19\% for the baseline BiLSTM. This confirms the advantage of parallel attention in capturing diverse phishing patterns that may be overlooked by simpler architectures.

\subsubsection{\textbf{Knowledge Distillation Results and Comparison with MobileBERT}}

Table \ref{Table_Results_Knowledge_Distillation} presents the average 5-fold cross-validation results for the BiLSTM model enhanced with multi-head attention and knowledge distillation from MobileBERT. Despite having 5.6 times fewer parameters (4.5M vs. 25.3M), the student model achieves performance comparable to its teacher. When trained on the combined dataset, the student model reaches an F1 score of 96.67\%, only 1.89\% lower than MobileBERT’s 98.56\%, while requiring less than half the training time (1,301.72s vs. 3,020s) and being approximately 7 times faster in inference (6.06s vs. 42s). 


The performance of KD-BiLSTM matches MobileBERT in all distribution scenarios as shown in Figure ~\ref{fig:eval_threeplots_composite}(c) which demonstrates its potential to replace transformer models with a lightweight approach.

\begin{figure*}[!t]
  \centering
  \setlength{\tabcolsep}{2pt}
  \renewcommand{\arraystretch}{0.9}

  \begin{tabular}{@{}c c@{}}
    \includegraphics[width=0.49\textwidth]{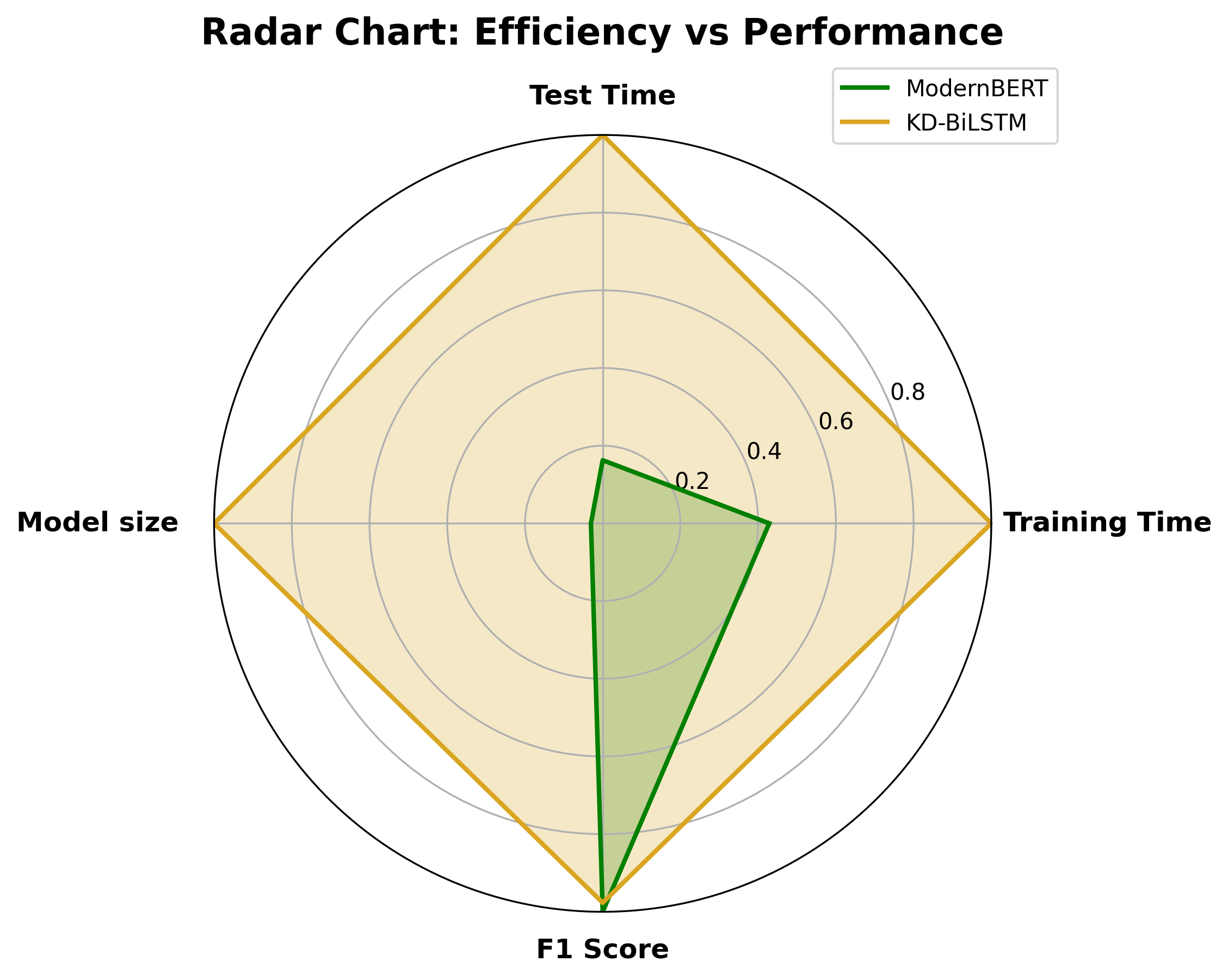} &
    \includegraphics[width=0.49\textwidth]{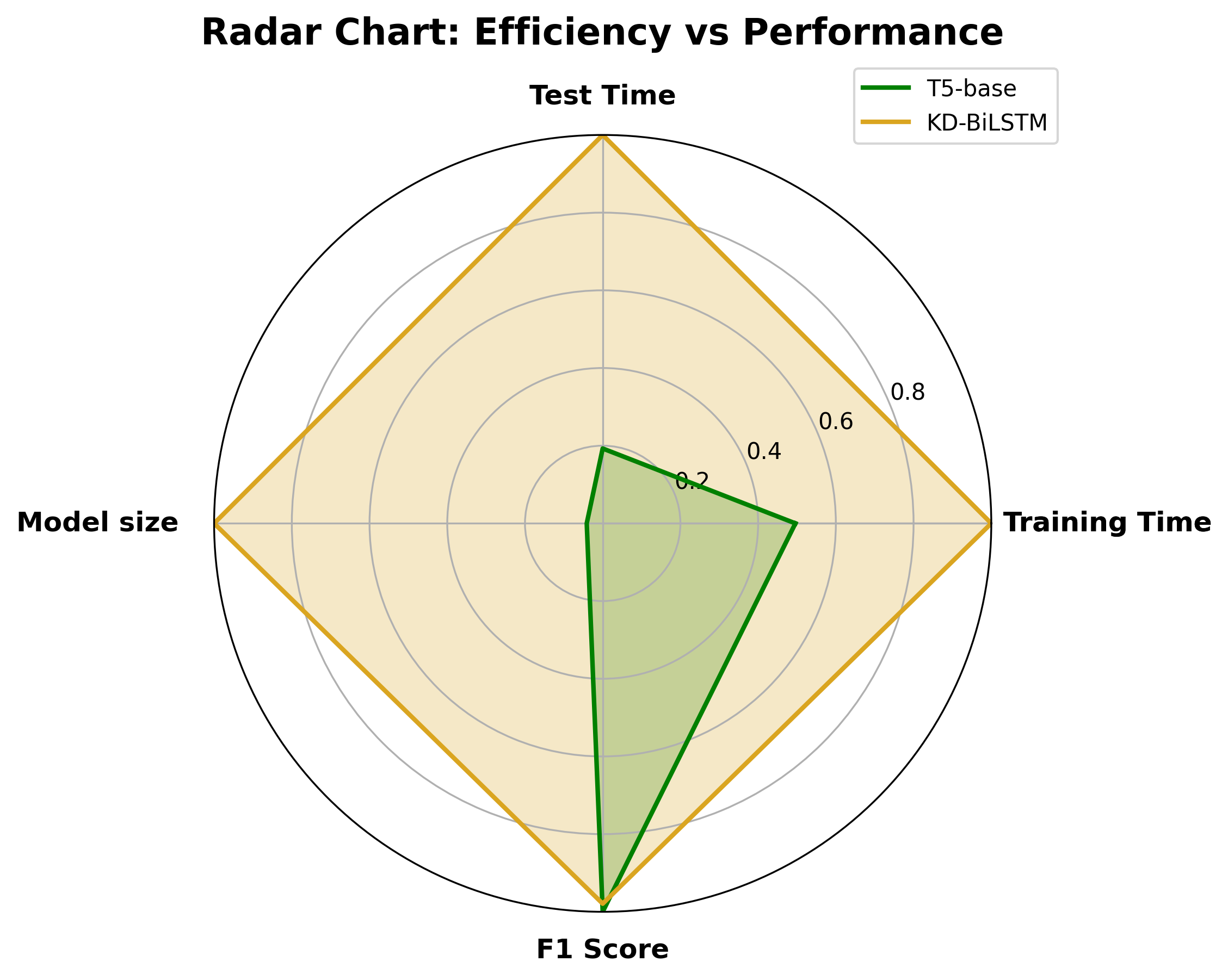} \\
    \footnotesize (a) KD-BiLSTM vs ModernBERT-base &
    \footnotesize (b) KD-BiLSTM vs T5-base \\[6pt]

    \includegraphics[width=0.49\textwidth]{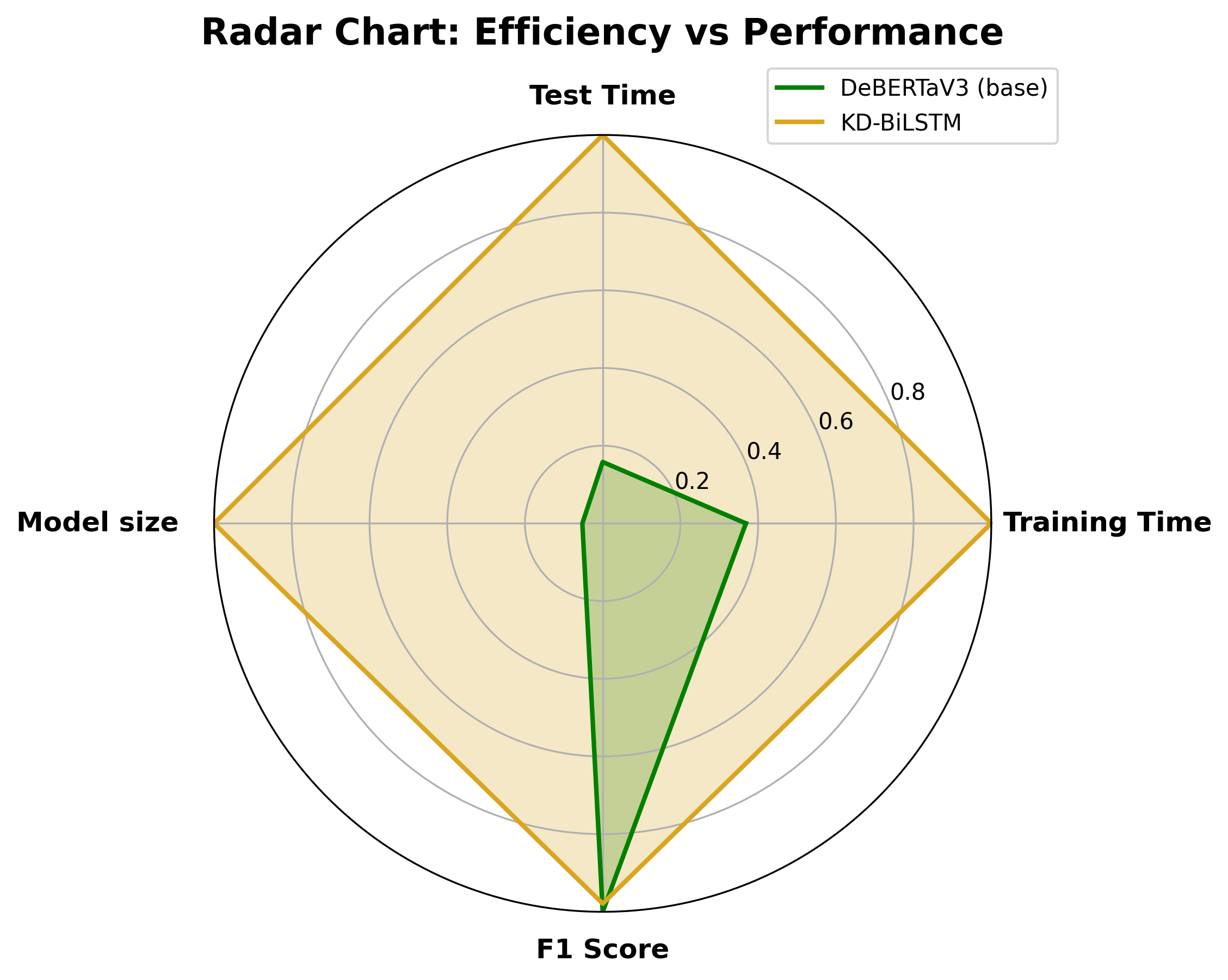} &
    \includegraphics[width=0.49\textwidth]{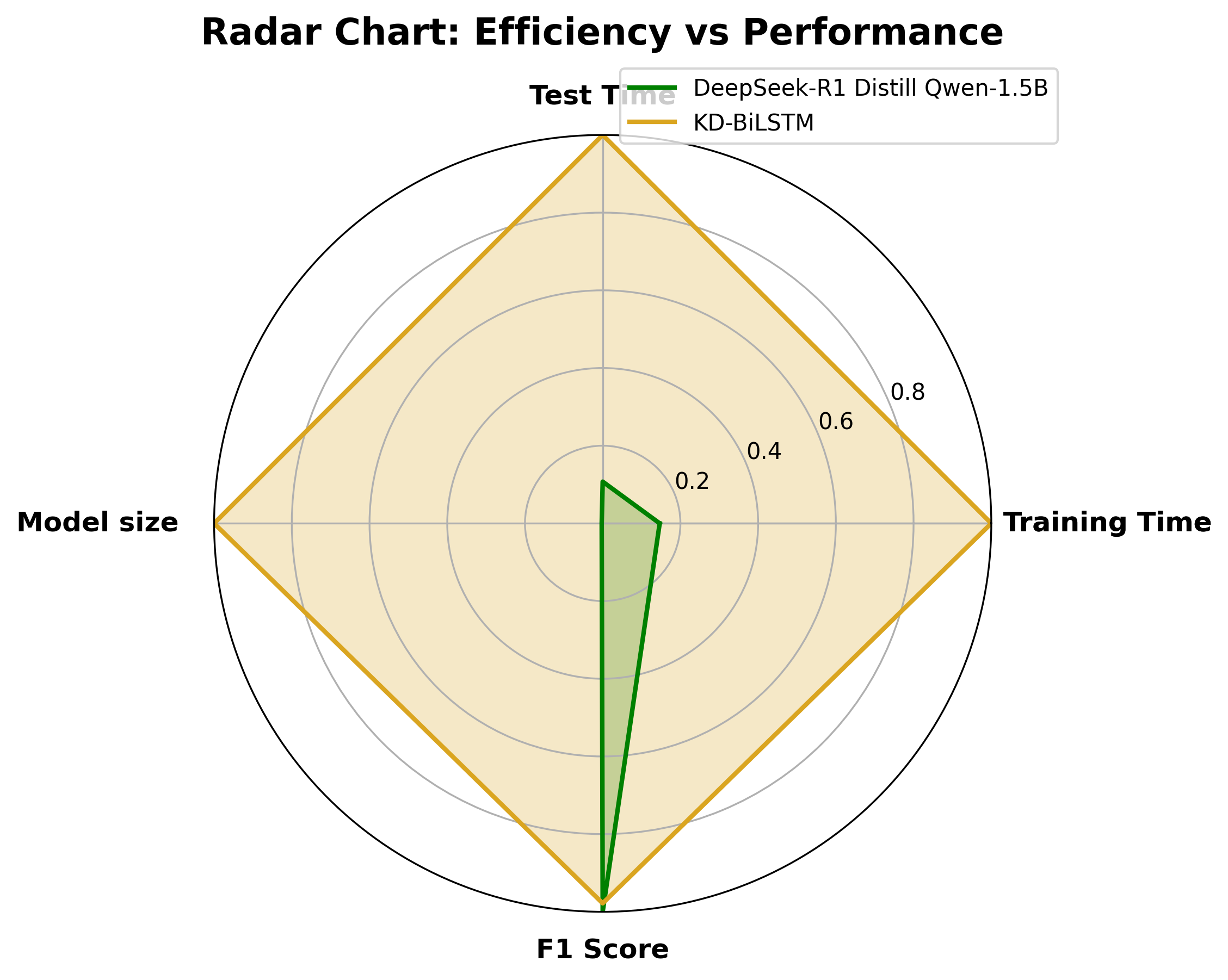} \\
    \footnotesize (c) KD-BiLSTM vs DeBERTaV3 (base) &
    \footnotesize (d) KD-BiLSTM vs DeepSeek-R1 Distill Qwen-1.5B \\[6pt]

    \includegraphics[width=0.49\textwidth]{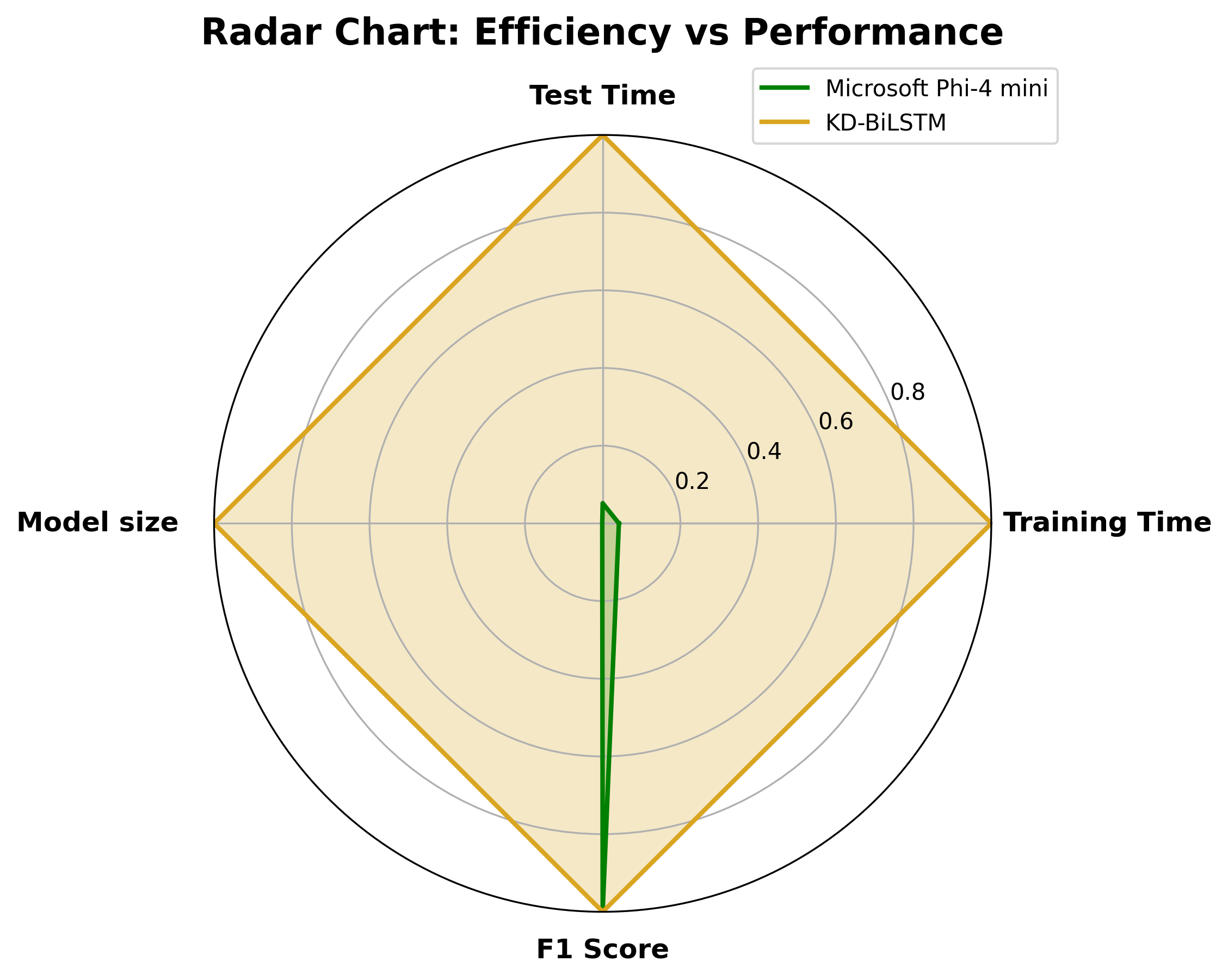} &
    \includegraphics[width=0.49\textwidth]{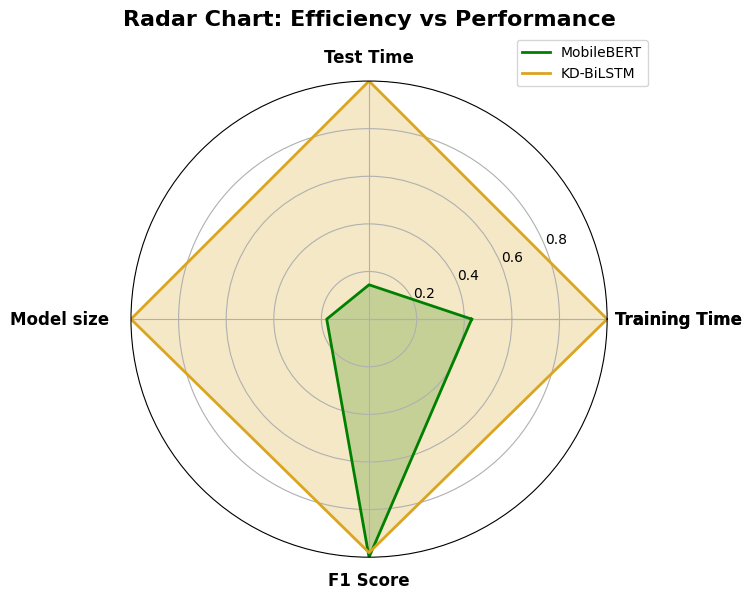} \\
    \footnotesize (e) KD-BiLSTM vs Microsoft Phi-4 mini (cls head) &
    \footnotesize (f) KD-BiLSTM vs MobileBERT (teacher) \\
  \end{tabular}

  \caption{Mixture-of-all split. Pareto-style radar comparisons of KD-BiLSTM vs six baselines. Axes are min–max normalized with time/size inverted (higher is better).}
  \label{fig:radar_grid_3x2}
\end{figure*}

This significant improvement in efficiency makes the student model highly suitable for resource-constrained deployments, such as email clients or edge devices. The Pareto view in Figure 11 illustrates these trade-offs by showing that all axes have a "larger is better" direction which demonstrates that KD-BiLSTM achieves better efficiency while maintaining competitive F1 scores compared to the baselines.
 As shown in the figure \ref{fig:radar_grid_3x2}, 
The student model operates with 4.5M parameters while delivering 5--19$\times$ faster inference performance and 20--800$\times$ smaller size compared to large Transformer baselines while maintaining F1 scores within 1--2.3 points. 
The KD-BiLSTM model achieves superior performance than multiple large models when evaluating Orig to Gen which demonstrates its ability to handle LLM-rewritten emails. 
The proposed model provides better suitability for gateway and on-device deployments because it achieves better latency and size performance even though some baselines maintain their top F1 scores.

These results confirm that knowledge distillation successfully transfers semantic generalization capabilities from MobileBERT to the lightweight BiLSTM model without sacrificing detection effectiveness. 

\textbf{Deployment implication}. The end-to-end cost profile (Test time, Params) determines the outcome for inline filtering and endpoint use even though absolute F1 reaches its peak on large transformers. The 4.5 M parameter size of KD-BiLSTM along with its 5–19 times faster inference speed compared to 86–3840 M parameter baselines allows for tighter tail-latency constraints and more dense multi-tenant batching without requiring dedicated accelerators which provides greater benefits than 2.3-point F1 improvement on offline benchmarks



%% file: conclusion.tex
\label{conclusion}
The research developed an LLM-resilient deception detection system for enterprise communication that distills knowledge from a fine-tuned MobileBERT teacher into a lightweight bidirectional LSTM with multi-head attention. The student model maintains the semantic decision boundaries of the teacher model while providing a smaller size and predictable latency which makes it deployable at gateways and endpoints under privacy, memory, and throughput constraints.

The evaluation process involved five training/testing protocols including cross-distribution settings that combine human-written and LLM-generated text in a hybrid dataset that included examples of paraphrasing and masking and personalization techniques. The attention-augmented BiLSTM model achieved weighted-F1 within approximately 1–2.5 points of strong transformer baselines in the "mixture-of-all" split while reducing both inference time and parameter count relative to six latency-sensitive baselines: ModernBERT and DeBERTaV3-base and T5-base and DeepSeek-R1 Distill Qwen-1.5B and Phi-4 mini and MobileBERT teacher. The research demonstrates an favorable trade-off between accuracy and latency and model size which fulfills the security needs of application-layer defense systems.

The student model’s small size and reliable inference behavior make it suitable for on-device or in-path classification that preserves user content and complements reputation and sandboxing signals. The system achieves near-teacher accuracy while maintaining a lower operational cost which enables organizations to implement policy-compliant, privacy-preserving defenses against social-engineering attacks that preserve user privacy.

%% file: Acknowledgments.tex
The authors express their gratitude to the anonymous reviewers for their valuable feedback. Additionally, they sincerely appreciate the support from the Cambridge Cybercrime Centre at the University of Cambridge for providing access to their phishing sample repositories.

%% file: References.bib
@article{nahmias2024prompted,
  title={Prompted contextual vectors for spear-phishing detection},
  author={Nahmias, Daniel and Engelberg, Gal and Klein, Dan and Shabtai, Asaf},
  journal={arXiv preprint arXiv:2402.08309},
  year={2024}
}

@article{guo2024x,
  title={X-Phishing-Writer: A Framework for Cross-Lingual Phishing Email Generation},
  author={Guo, Shih-Wei and Fan, Yao-Chung},
  journal={ACM Transactions on Asian and Low-Resource Language Information Processing},
  year={2024},
  publisher={ACM New York, NY}
}

@article{das2019automated,
  title={Automated email generation for targeted attacks using natural language},
  author={Das, Avisha and Verma, Rakesh},
  journal={arXiv preprint arXiv:1908.06893},
  year={2019}
}

@article{alkhalil2021phishing,
  title={Phishing attacks: A recent comprehensive study and a new anatomy},
  author={Alkhalil, Zainab and Hewage, Chaminda and Nawaf, Liqaa and Khan, Imtiaz},
  journal={Frontiers in Computer Science},
  volume={3},
  pages={563060},
  year={2021},
  publisher={Frontiers}
}

@article{thakur2023systematic,
  title={A systematic review on deep-learning-based phishing email detection},
  author={Thakur, Kutub and Ali, Md Liakat and Obaidat, Muath A and Kamruzzaman, Abu},
  journal={Electronics},
  volume={12},
  number={21},
  pages={4545},
  year={2023},
  publisher={MDPI}
}

@article{heiding2024devising,
  title={Devising and detecting phishing emails using large language models},
  author={Heiding, Fredrik and Schneier, Bruce and Vishwanath, Arun and Bernstein, Jeremy and Park, Peter S},
  journal={IEEE Access},
  year={2024},
  publisher={IEEE}
}

@article{koide2024chatspamdetector,
  title={Chatspamdetector: Leveraging large language models for effective phishing email detection},
  author={Koide, Takashi and Fukushi, Naoki and Nakano, Hiroki and Chiba, Daiki},
  journal={arXiv preprint arXiv:2402.18093},
  year={2024}
}

@article{hazell2023spear,
  title={Spear phishing with large language models},
  author={Hazell, Julian},
  journal={arXiv preprint arXiv:2305.06972},
  year={2023}
}

@misc{CambridgeCybercrimeCentre,
  author       = {{Cambridge Cybercrime Centre}},
  title        = {Cambridge Cybercrime Centre},
  howpublished = {\url{https://www.cambridgecybercrime.uk/}},
  note         = {Accessed: 2025-05-12},
  year         = {2015}
}

@misc{phishing_pot,
  author       = {Rodrigo F. Peixoto},
  title        = {phishing\_pot: A collection of phishing samples for researchers and detection developers},
  year         = {2025},
  url          = {https://github.com/rf-peixoto/phishing_pot},
  note         = {Accessed: 2025-05-02}
}

@misc{subhajournal_phishingemails,
  author       = {{Subhadeep Chakraborty}},
  title        = {Chakraborty Phishing Dataset},
  year         = {2023},
  howpublished = {\url{https://www.kaggle.com/datasets/subhajournal/phishingemails}},
  note         = {Accessed: 2025-05-02}
}

@misc{EnronEmailDataset,
  author       = {{William W. Cohen}},
  title        = {Enron Email Dataset},
  howpublished = {\url{https://www.cs.cmu.edu/~enron/}},
  note         = {Accessed: 2025-05-12},
  year         = {2015}
}

@article{hinton2015distilling,
  title={Distilling the knowledge in a neural network},
  author={Hinton, Geoffrey and Vinyals, Oriol and Dean, Jeff},
  journal={arXiv preprint arXiv:1503.02531},
  year={2015}
}

@inproceedings{wang2013study,
  title={A study on evolution of email spam over fifteen years},
  author={Wang, De and Irani, Danesh and Pu, Calton},
  booktitle={9th IEEE international conference on collaborative computing: networking, applications and worksharing},
  pages={1--10},
  year={2013},
  organization={IEEE}
}

@article{oard2015avocado,
  title={Avocado research email collection},
  author={Oard, Douglas and Webber, William and Kirsch, David and Golitsynskiy, Sergey},
  journal={Philadelphia: Linguistic Data Consortium},
  year={2015}
}

@misc{spamassassin_corpus,
  author       = {{Apache SpamAssassin Project}},
  title        = {SpamAssassin Public Mail Corpus},
  year         = {2006},
  howpublished = {\url{https://spamassassin.apache.org/old/publiccorpus/}},
  note         = {Accessed: 2025-05-28}
}

@article{kim2021comparing,
  title={Comparing kullback-leibler divergence and mean squared error loss in knowledge distillation},
  author={Kim, Taehyeon and Oh, Jaehoon and Kim, NakYil and Cho, Sangwook and Yun, Se-Young},
  journal={arXiv preprint arXiv:2105.08919},
  year={2021}
}

@article{wang2022tcurl,
  title={TCURL: Exploring hybrid transformer and convolutional neural network on phishing URL detection},
  author={Wang, Chenguang and Chen, Yuanyuan},
  journal={Knowledge-Based Systems},
  volume={258},
  pages={109955},
  year={2022},
  publisher={Elsevier}
}

@article{chiew2018survey,
  title={A survey of phishing attacks: Their types, vectors and technical approaches},
  author={Chiew, Kang Leng and Yong, Kelvin Sheng Chek and Tan, Choon Lin},
  journal={Expert Systems with Applications},
  volume={106},
  pages={1--20},
  year={2018},
  publisher={Elsevier}
}

@article{mikolov2013efficient,
  title={Efficient estimation of word representations in vector space},
  author={Mikolov, Tomas and Chen, Kai and Corrado, Greg and Dean, Jeffrey},
  journal={arXiv preprint arXiv:1301.3781},
  year={2013}
}

@inproceedings{gonzalez2011phishing,
  title={Phishing by form: The abuse of form sites},
  author={Gonzalez, Hugo and Nance, Kara and Nazario, Jose},
  booktitle={2011 6th International Conference on Malicious and Unwanted Software},
  pages={95--101},
  year={2011},
  organization={IEEE}
}

@article{gui2024psc,
  title={PSC-BERT: A spam identification and classification algorithm via prompt learning and spell check},
  author={Gui, Jiayi and Zhou, Yuhao and Yu, Ke and Wu, Xiaofei},
  journal={Knowledge-Based Systems},
  volume={301},
  pages={112266},
  year={2024},
  publisher={Elsevier}
}

@article{gou2021knowledge,
  title={Knowledge distillation: A survey},
  author={Gou, Jianping and Yu, Baosheng and Maybank, Stephen J and Tao, Dacheng},
  journal={International journal of computer vision},
  volume={129},
  number={6},
  pages={1789--1819},
  year={2021},
  publisher={Springer}
}

@article{wang2022generalizing,
  title={Generalizing to unseen domains: A survey on domain generalization},
  author={Wang, Jindong and Lan, Cuiling and Liu, Chang and Ouyang, Yidong and Qin, Tao and Lu, Wang and Chen, Yiqiang and Zeng, Wenjun and Yu, Philip S},
  journal={IEEE transactions on knowledge and data engineering},
  volume={35},
  number={8},
  pages={8052--8072},
  year={2022},
  publisher={IEEE}
}

@article{zhou2022domain,
  title={Domain generalization: A survey},
  author={Zhou, Kaiyang and Liu, Ziwei and Qiao, Yu and Xiang, Tao and Loy, Chen Change},
  journal={IEEE transactions on pattern analysis and machine intelligence},
  volume={45},
  number={4},
  pages={4396--4415},
  year={2022},
  publisher={IEEE}
}

@article{he2020deberta,
  title={Deberta: Decoding-enhanced bert with disentangled attention},
  author={He, Pengcheng and Liu, Xiaodong and Gao, Jianfeng and Chen, Weizhu},
  journal={arXiv preprint arXiv:2006.03654},
  year={2020}
}

@article{warner2024smarter,
  title={Smarter, better, faster, longer: A modern bidirectional encoder for fast, memory efficient, and long context finetuning and inference},
  author={Warner, Benjamin and Chaffin, Antoine and Clavi{\'e}, Benjamin and Weller, Orion and Hallstr{\"o}m, Oskar and Taghadouini, Said and Gallagher, Alexis and Biswas, Raja and Ladhak, Faisal and Aarsen, Tom and others},
  journal={arXiv preprint arXiv:2412.13663},
  year={2024}
}

@article{sun2020mobilebert,
  title={Mobilebert: a compact task-agnostic bert for resource-limited devices},
  author={Sun, Zhiqing and Yu, Hongkun and Song, Xiaodan and Liu, Renjie and Yang, Yiming and Zhou, Denny},
  journal={arXiv preprint arXiv:2004.02984},
  year={2020}
}

@article{guo2025deepseek,
  title={Deepseek-r1: Incentivizing reasoning capability in llms via reinforcement learning},
  author={Guo, Daya and Yang, Dejian and Zhang, Haowei and Song, Junxiao and Zhang, Ruoyu and Xu, Runxin and Zhu, Qihao and Ma, Shirong and Wang, Peiyi and Bi, Xiao and others},
  journal={arXiv preprint arXiv:2501.12948},
  year={2025}
}

@article{nasution2025benchmarking,
  title={Benchmarking 21 Open-Source Large Language Models for Phishing Link Detection with Prompt Engineering},
  author={Nasution, Arbi Haza and Monika, Winda and Onan, Aytug and Murakami, Yohei},
  journal={Information},
  volume={16},
  number={5},
  pages={366},
  year={2025},
  publisher={MDPI}
}

@article{motlagh2024large,
  title={Large language models in cybersecurity: State-of-the-art},
  author={Motlagh, Farzad Nourmohammadzadeh and Hajizadeh, Mehrdad and Majd, Mehryar and Najafi, Pejman and Cheng, Feng and Meinel, Christoph},
  journal={arXiv preprint arXiv:2402.00891},
  year={2024}
}

@misc{ENISA2023ThreatLandscape,
  author    = {{ENISA}},
  title     = {ENISA Threat Landscape 2023},
  howpublished = {\url{https://www.enisa.europa.eu/}},
  year      = {2023},
  note      = {European Union Agency for Cybersecurity annual report}
}

@article{ebrahimi2017hotflip,
  title={Hotflip: White-box adversarial examples for text classification},
  author={Ebrahimi, Javid and Rao, Anyi and Lowd, Daniel and Dou, Dejing},
  journal={arXiv preprint arXiv:1712.06751},
  year={2017}
}

@article{he2021debertav3,
  title={Debertav3: Improving deberta using electra-style pre-training with gradient-disentangled embedding sharing},
  author={He, Pengcheng and Gao, Jianfeng and Chen, Weizhu},
  journal={arXiv preprint arXiv:2111.09543},
  year={2021}
}

@article{raffel2020exploring,
  title={Exploring the limits of transfer learning with a unified text-to-text transformer},
  author={Raffel, Colin and Shazeer, Noam and Roberts, Adam and Lee, Katherine and Narang, Sharan and Matena, Michael and Zhou, Yanqi and Li, Wei and Liu, Peter J},
  journal={Journal of machine learning research},
  volume={21},
  number={140},
  pages={1--67},
  year={2020}
}

@article{abouelenin2025phi,
  title={Phi-4-mini technical report: Compact yet powerful multimodal language models via mixture-of-loras},
  author={Abouelenin, Abdelrahman and Ashfaq, Atabak and Atkinson, Adam and Awadalla, Hany and Bach, Nguyen and Bao, Jianmin and Benhaim, Alon and Cai, Martin and Chaudhary, Vishrav and Chen, Congcong and others},
  journal={arXiv preprint arXiv:2503.01743},
  year={2025}
}
